\documentclass[12pt,a4paper]{iopart}

\usepackage{array}
\usepackage{graphicx}
\usepackage{dcolumn}
\usepackage{bm}
\usepackage{upgreek}
\usepackage{amssymb}
\usepackage{url}
\usepackage{xcolor}
\usepackage{MnSymbol}

\newcommand{\I}{\mathrm{i}}
\newcommand{\E}{\mathrm{e}}
\newcommand{\abs}{\mathrm{abs}}
\newcommand{\emi}{\mathrm{emi}}

\newcommand{\XUV}{\mathrm{XUV}}

\newcommand{\asym}{\mathrm{asym}}

\newcommand{\oneph}{\mathrm{(1ph)}}
\newcommand{\twoph}{\mathrm{(2ph)}}
\newcommand{\CC}{\mathrm{CC}}
\newcommand{\dd}{\mathrm{d}}

\begin{document}

\title[Analytical Continuum-continuum Transition Amplitude of H]{Analytical Expression for Continuum-continuum Transition Amplitude of Hydrogen-like Atoms with Angular-momentum Dependence}

\author{J B Ji$^{1,*}$, 
        K Ueda$^{1,2,3,*}$,
        M Han$^{1,4}$
        and
        H J W{\"o}rner$^{1,*}$}
\address{$^1$ Laboratorium f{\"u}r Physikalische Chemie, ETH Z{\"u}rich, 8093 Zürich, Switzerland}
\address{$^2$ Department of Chemistry, Tohoku University, Sendai, 980-8578, Japan}
\address{$^3$ School Physical Science and Technology, ShanghaiTech University, Shanghai 201210, China}
\address{$^4$ J. R. Macdonald Laboratory, Department of Physics, Kansas State University, Manhattan, KS 66506, USA}
\address{$^*$ Authors to whom any correspondence should be addressed.}
\ead{
\mailto{jiabao.ji@phys.chem.ethz.ch},
\mailto{kiyoshi.ueda@tohoku.ac.jp},
\mailto{hwoerner@ethz.ch}
}


\begin{abstract}
Attosecond chronoscopy typically utilises interfering two-photon transitions to access the phase information. Simulating these two-photon transitions is challenging due to the continuum-continuum transition term. The hydrogenic approximation within second-order perturbation theory has been widely used due to the existence of analytical expressions of the wave functions. So far, only (partially) asymptotic results have been derived, which fail to correctly describe the low-kinetic-energy behaviour, especially for high angular-momentum states. 
Here, we report an analytical expression that overcome these limitations. They are based on the Appell's $F_1$ function and use the confluent hypergeometric function of the second kind as the intermediate states. 
We show that the derived formula quantitatively agrees with the numerical simulations using the time-dependent Schr{\"o}dinger equation for various angular-momentum states, which improves the accuracy compared to the other analytical approaches that were previously reported.
Furthermore, we give an angular-momentum-dependent asymptotic form of the outgoing wavefunction and their continuum-continuum dipole transition amplitudes.
\end{abstract}

\maketitle


\section{Introduction}

\begin{figure}[bt]
\begin{center}
\includegraphics[width=0.6\textwidth]{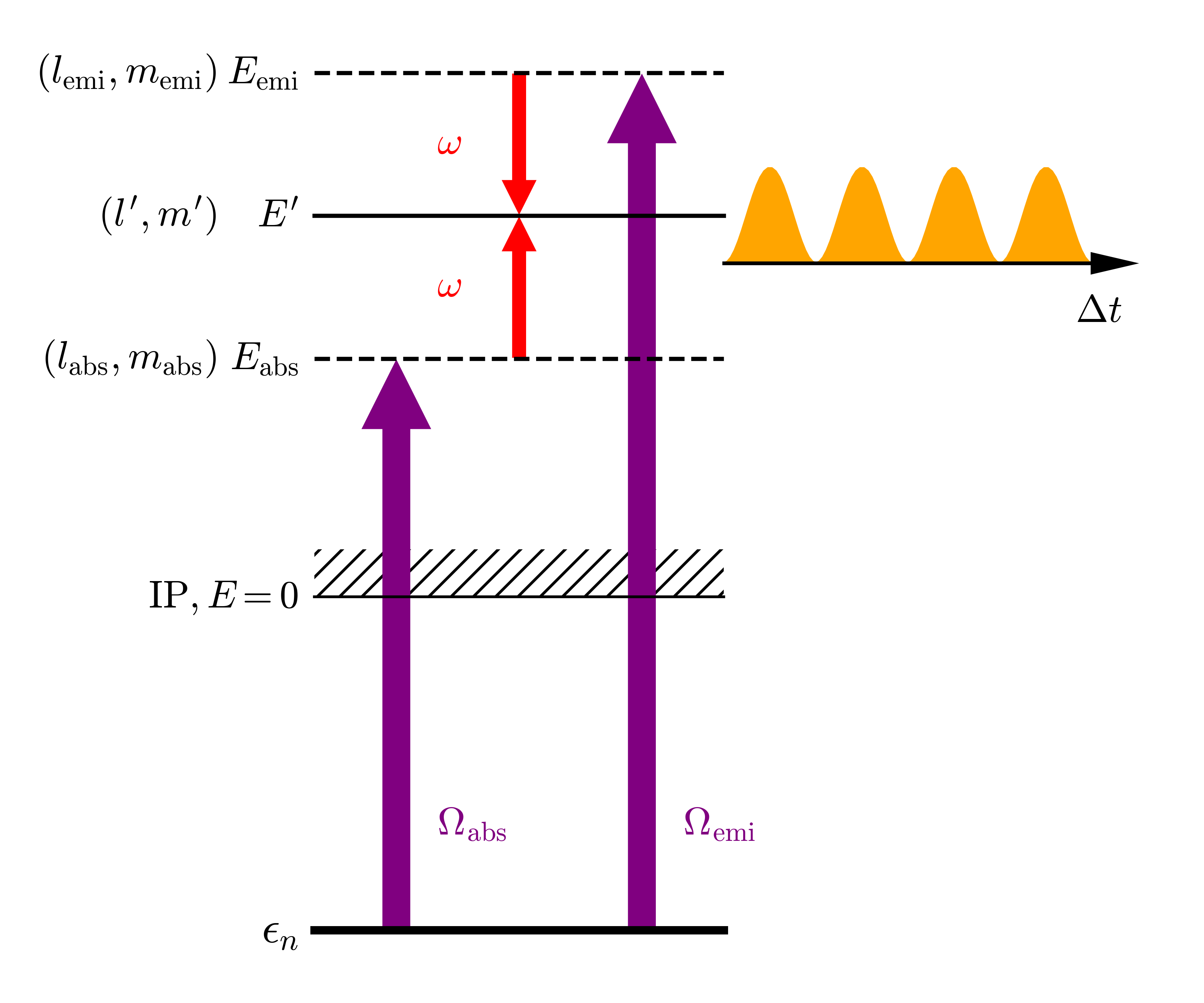}
\caption{Typical energy diagram of the RABBIT scheme. The electron is excited from the bound initial state with $\epsilon_n < 0$ with two frequency components $\Omega_{\abs}$ and $\Omega_{\emi}$ of the attosecond pulse train (purple arrows) to the corresponding intermediate states (dashed lines) with electron kinetic energies of $E_{\abs}$ and $E_{\emi}$, respectively. The presence of the dressing field (red arrows) with frequency $\omega$ induces the CC transition from the intermediate states to the final state (sideband) with electron kinetic energy of $E'$ and leads to the oscillation of the intensity of the sideband as a function of the XUV-IR time delay $\Delta t$, as sketched by the orange wave. The partial-wave indices are denoted as $(l,m)$ for all states in the continuum, and the total transition amplitude is the superposition of the contributions from all possible pathways. \label{fig:RABBITscheme}}
\end{center}
\end{figure}

The question ``How long does the photoionisation take place?'' has intrigued researchers for decades ever since the early years of quantum theories. Its natural timescale of attosecond (as, ${10}^{-18}$ s) has introduced a new field of science, known as the attosecond science, which is highlighted by the 2023 Nobel Prize in Physics \cite{nobelprizeNobelPrize}. The last 30 years have witnessed the rapid development of the toolbox for investigating the photoionisation processes with phase information embedded. This includes, but is not limited to, high-harmonic generation (HHG) \cite{mcpherson87a,ferray88a} as a table-top XUV source thanks to the availability of the high-power lasers, the schemes of performing the phase measurement such as attosecond streaking \cite{itatani02a,kienberger03a,gaumnitz17a} and the reconstruction of attosecond beating by interference of two-photon transitions (RABBIT) \cite{paul01a,muller02a}, and three-dimensional momentum-resolved particle detectors such as the cold target recoil ion momentum spectrometer (COLTRIMS) \cite{dorner00a,ullrich03a,gong2022attosecond}, and so on. As a result, it has become possible to measure the Wigner time delay defined as \cite{eisenbud1948formal,wigner55a,smith60a}: 
\begin{equation}
    \tau(E) = \hbar \frac{\partial \arg \{ \mathfrak{T}(E) \}}{\partial E}
    \label{eq:tau_Wigner}
\end{equation}
where $\mathfrak{T}$ is the transition amplitude for photoionisation, and in the following discussions in this article, unless otherwise stated, we use the atomic units, where $e = m_e = \hbar = 1$, and the atomic unit of time $1~{\rm a.u.} \approx 2.42 \times {10}^{-17} ~ {\rm s} = 24.2~ {\rm as}$. 
Although this definition is rather straightforward, and the transition amplitudes (phases and moduli) of one-photon ionisation for atoms and molecules can be routinely calculated by e.g. random-phase approximation with exchange (RPAE) \cite{amusia1972interference,kheifets13a}, time-dependent local-density approximation (TDLDA) \cite{zangwill1980density,dixit2013time} and Schwinger variational methods such as \texttt{ePolyScat} \cite{gianturco94a,natalense99a}, in reality, the schemes of the phase measurements involve two-photon transitions, which differ from one-photon ionisations. 
An illustration of the RABBIT measurement scheme is illustrated in figure \ref{fig:RABBITscheme}. The attosecond pulse train (APT) created by HHG contains typically odd harmonics of the fundamental driving field ($\omega$), which leads to the main bands (dashed lines in figure \ref{fig:RABBITscheme}). If a copy of the driving field is spatially and temporally overlapped with the APT as the dressing field, an additional sideband (solid line in figure \ref{fig:RABBITscheme}) lying between the two main bands appears, which corresponds to either absorbing one photon from the lower main band (the absorption pathway), or emitting one photon from the higher main band (the emission pathway). 
The transition amplitudes of the absorption and the emission pathways can be written as:
\begin{equation}
    \mathcal{A}_{\mp}(\Delta t) 
    = \mathcal{A}_{\mp,0} \exp(\pm \I \omega \Delta t)
\end{equation}
where $(-)$ and $(+)$ in the subscript are for the absorption and the emission pathways, respectively. The relative time delay of the dressing field is denoted by $\Delta t$ (the dressing field has the light path $c \Delta t$ longer than the attosecond pulse, where $c$ is the speed of light), meaning that at the time of the attosecond pulse burst the dressing field has the phase of $\omega \Delta t$. The transition amplitudes at $\Delta t = 0$ are denoted as $\mathcal{A}_{\mp,0}$. The phase of the dressing field contributes to the absorption and the emission pathways oppositely, and the side-band intensity is expressed by:
\begin{eqnarray}
    \fl \mathcal{S}(\Delta t)
    = {\left| \mathcal{A}_{\abs} \E^{\I \omega \Delta t} + \mathcal{A}_{\emi} \E^{-\I \omega \Delta t}\right|}^2
    = {\left| \mathcal{A}_{\abs} \right|}^2
    + {\left| \mathcal{A}_{\emi} \right|}^2
    + 2 \left| \mathcal{A}_{\abs} \mathcal{A}_{\emi} \right|
    \cos (2 \omega \Delta t - \phi)
    \label{eq:RABBIT_SB_intensity_Delta_t}
\end{eqnarray}
where $\phi = \arg \{ \mathcal{A}_{\emi} \} - \arg \{ \mathcal{A}_{\abs} \}$ is the phase difference between the emission and the absorption pathways. We can further write
\begin{equation}
    \frac{\phi}{2 \omega} = \tau_{\XUV} + \tau_{\rm A}
\end{equation}
where $\tau_{\XUV} = (\phi^{\XUV}_{+} - \phi^{\XUV}_{-})/(2 \omega)$ corresponds to the relative phase between the harmonic frequencies in the attosecond pulse train, also known as the attochirp, and $\tau_{\rm A} = (\arg \{ \mathfrak{T}^{\twoph}_{\emi} \} -\arg \{ \mathfrak{T}^{\twoph}_{\abs} \})/(2 \omega)$ is known as the two-photon ionisation time delay, since it corresponds to the phase difference of the two-photon transition amplitudes between the emission and the absorption pathways. More discussions and examples can be found in the book chapter \cite{Busto2024Atomic}.

From the theoretical point of view, the \textit{ab initio} computation such as numerically solving the time-dependent Schr{\"o}dinger equation (TDSE) is limited to the modelled potential with one active electron, e.g. the hydrogen-like atoms, while modelling the potential for molecules can be very challenging. 
A common approach for handling this problem is to separate the two-photon transition into two parts, i.e. the bound-continuum transition and the continuum-continuum (CC) transition, so that the two parts can be treated individually. 
This essentially stems from the lowest-order perturbation theory. 
The two-photon transition matrix element from an bound initial state $| n \rangle$ to a final state in the continuum with momentum $\vec{k}$ can be represented as \cite{dahlstrom13a}:
\begin{equation}
    \mathcal{M}_n(\vec{k}';\Omega)
    = \frac{1}{\I} 
    \lim_{\varepsilon \rightarrow 0^+}
    \sumint_{\nu}
    \frac{{\langle \vec{k}' | \hat{d}_{\omega} | \nu \rangle}{\langle \nu | \hat{d}_{\Omega} | n \rangle}}{\epsilon_n + \Omega - \epsilon_{\nu} + \I \varepsilon}
    \label{eq:M_two_photon}
\end{equation}
and the two-photon transition amplitude reads
\begin{equation}
    \mathfrak{T}^{\twoph}_{n \rightarrow l',m'}(E';\theta,\varphi)
    = \mathcal{E}_{\omega} \mathcal{E}_{\Omega}
    \mathcal{M}_n(\vec{k}';\Omega)
\end{equation}
where $\mathcal{E}_{\Omega}$ and $\mathcal{E}_{\omega}$ are the complex amplitudes (field strengths with phases included) of the XUV and dressing fields, respectively, and $\vec{k}'$ is the vector with modulus of $\sqrt{2E'}$ and pointing towards the emission angle $(\theta,\varphi)$, either in the lab frame or in the molecular frame for molecules. Within the framework of dipole approximation, the dipole terms for the XUV and dressing fields are $\hat{d}_{\Omega}$ and $\hat{d}_{\omega}$, respectively. The $\sumint_{\nu}$ sign refers to summing over all discrete states and integrating over all continuous states. 
If one assumes that the two-photon process is dominated by the pathway that the electron absorbs one XUV photon from the initial state to an intermediate state in the continuum (assuming that the XUV photon energy is higher than the photoionisation energy) with the corresponding kinetic energy, after which a CC transition occurs due to the presence of the dressing IR field, where the electron either absorbs or emits one IR photon (multi-IR processes are ignored here, assuming the dressing field is relatively weak), as sketched in figure \ref{fig:RABBITscheme}, then the two-photon transition amplitude can be expressed as:
\begin{equation}
    \mathfrak{T}^{\twoph}_{n \rightarrow l',m'}(E') = \sum_{l,m} 
    \mathfrak{T}^{\oneph}_{n \rightarrow l,m}(E) 
    \mathfrak{T}^{(\CC)}_{l,m \rightarrow l',m'}(E, E')
    \label{eq:T_2ph_T_1ph_T_CC}
\end{equation}
where the terms $\mathfrak{T}^{\oneph}(E)$ and $\mathfrak{T}^{(\CC)}(E, E')$ refer to the bound-continuum and continuum-continuum transitions, respectively. Here we have decomposed the emission-angle-dependent transition amplitude into the spherical harmonics, i.e.
\begin{equation}
    \mathfrak{T}^{\twoph}_{n}(E';\theta,\varphi)
    = \sum_{l',m'} 
    \mathfrak{T}^{\twoph}_{n \rightarrow l',m'}(E')
    Y_{l'}^{m'}(\theta,\varphi)
    \label{eq:T_2ph_angle_lm}
\end{equation}
where $Y_{l'}^{m'}(\theta,\varphi)$ is the spherical harmonic (in the Condon–Shortley phase convention) with angular momentum and magnetic quantum numbers of $l'$ and $m'$, respectively. 
For the continuum-continuum part, the transition amplitude can be expressed as the product of the angular part and the radial part: 
\begin{eqnarray}
    \fl \mathfrak{T}^{(\CC)}_{l,m \rightarrow l',m'}(E, E')
    = \mathcal{E}_{\omega}
    \int_{4\pi} {\left(\ Y_{l'}^{m'} \right)}^* Y_1^{\mu} Y_l^m \dd \Omega 
    \times (-\pi) \int_0^\infty {\left( R_{l'}(E) \right)}^* r R_l(E) r^2 \dd r
    \label{eq:T_CC_anglular_radial}
\end{eqnarray}
where $Y_1^{\mu}$ is the polarisation of the dressing photon represented in the spherical harmonics. The $(-\pi)$-factor originates from the contour integral around the pole in equation (\ref{eq:M_two_photon}), which is explained in detail in \cite{dahlstrom13a}. The angular part can be calculated by the Clebsch–Gordan coefficients: 
\begin{equation}
    \fl \int_{4\pi} {\left(\ Y_{l'}^{m'} \right)}^* Y_1^{\mu} Y_l^m \dd \Omega
    = \sqrt{\frac{3 (2 l + 1)}{4 \pi (2 l' + 1)}}
    \langle l, 0, 1,   0  | l', 0  \rangle
    \langle l, m, 1, \mu  | l', m' \rangle
\end{equation}
where $\langle j_1, m_1, j_2, m_2  | J, M \rangle$ is the Clebsch–Gordan coefficient and is real under the Condon–Shortley phase convention. 
The latter integral between the radial wavefunctions, with the $(-\pi)$-term included, is denoted as $\mathcal{T}_{l \rightarrow l'}(k,k')$ in the following discussion, where $k = \sqrt{2E}$ and $k' = \sqrt{2E'}$ are the momenta of the intermediate and final states, respectively. 
The radial wavefunctions in principle depend on the shape of the radial potential, which can be complicated for molecular cases. If we assume that the CC transition is dominated by the long-range effect and is not sensitive to the local potential, then we can use the transition amplitudes derived from the hydrogen-like atoms. 
Therefore, instead of expression (\ref{eq:tau_Wigner}), the two-photon Wigner-like time delay is defined using the finite-difference method: 
\begin{equation}
    \tau^{\twoph}(E') = \frac{1}{2\omega}
    \left[ \arg \{ \mathfrak{T}^{\twoph}_{\emi}(E') \} - \arg \{ \mathfrak{T}^{\twoph}_{\abs}(E') \} \right]
    \label{eq:T_2ph_finite_difference}
\end{equation}
which can be represented either in the spherical-coordinate basis $(\theta,\varphi)$ or in the spherical-harmonic $(l,m)$ basis. If we further assume that the CC transition amplitude in equation (\ref{eq:T_2ph_T_1ph_T_CC}) does not depend on $(l,m)$ or $(l',m')$, then the two-photon time delay can be separated into two parts: 
\begin{eqnarray}
    \fl \tau^{\twoph}(E') = \frac{1}{2\omega}
    \big[ \arg \{ \mathfrak{T}^{\oneph}(E_{\emi}) \mathfrak{T}^{(\CC)}(E_{\emi}, E') \} 
    - \arg \{ \mathfrak{T}^{\oneph}(E_{\abs}) \mathfrak{T}^{(\CC)}(E_{\abs}, E') \} \big]
    \nonumber \\
    =  \frac{1}{2\omega} 
    \Big \{
    \left[ \arg \{ \mathfrak{T}^{\oneph}(E_{\emi}) \} - \arg \{ \mathfrak{T}^{\oneph}(E_{\abs}) \} \right]
    \nonumber \\
    + \left[ \arg \{  \mathfrak{T}^{(\CC)}(E_{\emi}, E') \} - \arg \{  \mathfrak{T}^{(\CC)}(E_{\abs}, E') \} \right]
    \Big \} \nonumber \\
    \approx \tau^{\oneph}(E') + \tau_{\CC}(E';\omega)
    \label{eq:tau_2ph_separate}
\end{eqnarray}
where we have approximated the derivative in expression (\ref{eq:tau_Wigner}) by its finite difference, and the CC time delay is defined as 
\begin{eqnarray}
    \tau_{\CC}(E';\omega) = \frac{1}{2\omega} 
    \left[ \arg \{  \mathfrak{T}^{(\CC)}(E_{\emi}, E') \} - \arg \{  \mathfrak{T}^{(\CC)}(E_{\abs}, E') \} \right] .
\end{eqnarray} 
By separating the two-photon time delay into the one-photon part and the continuum-continuum part, the problem is thus simplified as how to calculate the term $\mathfrak{T}^{(\CC)}(E,E')$, or equivalently the integral $\mathcal{T}(k,k')$ for the hydrogen-like atoms. Dahlstr{\"o}m \textit{et al.} reported two models based on the Wentzel–Kramers–Brillouin (WKB) approximation with asymptotic expansion, i.e. considering the phase modulation (P) and considering the phase and amplitude (modulus) modulation (P+A), both expressed by the analytical formulae involving the gamma function \cite{dahlstrom12a}: 
\begin{eqnarray}
    \fl \mathcal{T}^{(\mathrm{P})}_{\asym}(k,k')
    = \frac{\pi}{2} \frac{N_{k} N_{k'}}{{|k - k'|}^2}
    \exp \left[ -\frac{\pi Z}{2} (\frac{1}{k} - \frac{1}{k'}) \right]
    \frac{{(2k)}^{\I Z/k}}{{(2k')}^{\I Z/k'}}
    \frac{\Gamma[2 + \I Z (1/k - 1/k')]}{{(k-k')}^{\I Z (1/k - 1/k')}}
    \label{eq:T_CC_iso_P}
\end{eqnarray}
and 
\begin{eqnarray}
    \fl \mathcal{T}^{(\mathrm{P+A})}_{\asym}(k,k')
    = \mathcal{T}^{(\mathrm{P})}_{\asym}(k,k')
    \left[ 1 + \frac{\I Z}{2} \left( \frac{1}{k^2} + \frac{1}{{k'}^2} \right) \frac{k - k'}{1 + \I Z (1/k + 1/k')}  \right]
    \label{eq:T_CC_iso_PA}
\end{eqnarray}
where $N_{k} = \sqrt{2/(\pi k)}$ and $N_{k'} = \sqrt{2/(\pi k')}$ are the normalisation factors for the intermediate state and the final state, respectively. 
Note that the normalisation factor depends on normalising according to the cross section or normalising according to the flux. For the scattered wave the total flux is 
\begin{equation}
    \int_{4\pi} j_{\rm s}~r^2 ~ \dd \Omega = k \int_{4\pi} {|f_{\rm s}|}^2 ~ \dd \Omega = k \int_{4\pi} \frac{\dd \sigma}{\dd \Omega} ~ \dd \Omega
\end{equation}
where $j_{\rm s}$ is the scattering flux density, $f_{\rm s}$ is the scattering amplitude, and $\dd \sigma / \dd \Omega$ is the differential cross section. 
The well-known solutions to the Coulomb wave equation are conventionally normalised according to the cross section, while the WKB approximation yields the flux-based normalisation factor, which contains an additional factor $\propto k^{-1/2}$, which is clear from comparing their asymptotic oscillation moduli \cite{marante2014hybrid}. 
Regarding the phase and the corresponding time delay of the CC transitions, the TDSE results lie typically between these two models, and they converge towards the same values for high kinetic energies \cite{dahlstrom12a}. 
A number of works have explicitly \cite{gong2022attosecond,kluender11a,guenot12a,huppert16a,baykusheva17a,isinger17a,heck2021attosecond} or implicitly (e.g. to assume the two-photon time-delay difference between two species with the same or similar electron kinetic energies is equal or almost equal to their one-photon time-delay difference) \cite{jordan20a,heck2022two} used this assumption and have shown good agreement with the experimental results on molecules, indicating that this approximation is reasonable. 

Despite the great success of this elegant model in the past decade, some features of experimental observations cannot be explained by these models. For example, the phases of these models do not depend on the angular momentum $l$ of the intermediate or the final state, which causes that the angle-dependent time delay of the two-photon ionisation is simply that of the one-photon ionisation plus an isotropic CC time delay.
This can be proven as follows. Using equations (\ref{eq:T_2ph_T_1ph_T_CC}) and (\ref{eq:T_CC_anglular_radial}), we find that
\begin{eqnarray}
    \fl \mathfrak{T}^{\twoph}_{n \rightarrow l',m'}(E') = 
    \mathcal{E}_{\omega}
    \sum_{l,m} \mathfrak{T}^{\oneph}_{n \rightarrow l,m}(E)
    \mathcal{T}_{\asym}(\sqrt{2E},\sqrt{2E'})
    \int_{4\pi} {\left(\ Y_{l'}^{m'} \right)}^* Y_1^{\mu} Y_l^m \dd \Omega .
\end{eqnarray}
Writing this formula into the spherical coordinates using equation (\ref{eq:T_2ph_angle_lm}), we have
\begin{eqnarray}
    \fl \mathfrak{T}^{\twoph}_{n}(E';\theta',\varphi')
    = \mathcal{E}_{\omega}
    \mathcal{T}_{\asym}(\sqrt{2E},\sqrt{2E'})
    \int_{4\pi}
    \sum_{l',m'} Y_{l'}^{m'}(\theta',\varphi')
    \sum_{l,m} \mathfrak{T}^{\oneph}_{n}(E;\theta,\varphi)
    {\left( Y_l^m(\theta,\varphi) \right)}^*
    \nonumber \nonumber \\
    \times
    {\left(\ Y_{l'}^{m'}(\Tilde{\theta},\Tilde{\varphi}) \right)}^* Y_1^{\mu}(\Tilde{\theta},\Tilde{\varphi}) Y_l^m(\Tilde{\theta},\Tilde{\varphi}) ~ \dd \Tilde{\Omega}
    \nonumber \\
    = \mathcal{E}_{\omega}
    \mathcal{T}_{\asym}(\sqrt{2E},\sqrt{2E'})
    \int_{4\pi}
    \delta(\theta',\varphi';\Tilde{\theta},\Tilde{\varphi})
    \delta(\theta,\varphi;\Tilde{\theta},\Tilde{\varphi})
    \nonumber \\
    \quad \times
    Y_1^{\mu}(\Tilde{\theta},\Tilde{\varphi})
    \mathfrak{T}^{\oneph}_{n}(E;\theta,\varphi) ~ 
    \dd \Tilde{\Omega}
    \nonumber \\
    = \mathcal{E}_{\omega}
    \mathcal{T}_{\asym}(\sqrt{2E},\sqrt{2E'})
    Y_1^{\mu}(\theta',\varphi')
    \mathfrak{T}^{\oneph}_{n}(E;\theta',\varphi')
\end{eqnarray}
where the term $Y_1^{\mu}(\theta',\varphi')$ depends on the polarisation of the dressing field and does not contribute to the energy-dependent phase variation, i.e. the time delay. 
Hence, for helium one would expect an isotropic two-photon time delay, since its one-photon ionisation time delay has no angular dependence. This contradicts the experimental observation \cite{heuser16a,fuchs2020time,jiang2022atomic}, and the mechanism was demonstrated in \cite{busto2019fano} that the moduli of the transition amplitudes for $l \rightarrow l-1$ and $l \rightarrow l+1$ have preference depending on the increase or decrease of the electron kinetic energy, respectively, which is known as the Fano's propensity rule in the CC transition. The same mechanism was also confirmed by a more recent experimental work from the authors \cite{han2024separation} using the circular-XUV-circular-IR RABBIT, where the Fano's propensity rule is embodied by the angle-dependent RABBIT phase difference between the co- and counter-rotating XUV and IR. 

The missing piece was partially filled by Boll \textit{et al.} who replaced the final-state wavefunction from the asymptotic formula by the regular solution of the Coulomb wave equation, which is expressed by the Kummer's function, or the confluent hypergeometric function of the first kind ${_1F_1}$. The result can also be written as an analytical expression using Gauss' hypergeometric function ${_2F_1}$ \cite{boll2022analytical}. This model quantitatively reproduces Fano's propensity rule at high electron kinetic energies and the angle-resolved time delay of helium \cite{boll2022analytical,boll2023two}. However, the agreement deteriorates at lower energies (typically below 10 eV), particularly for larger $l'$s, and in the same energy range the isotropic formulae proposed by Dahlstr{\"o}m \textit{et al.} also show large discrepancy compared to the TDSE. On the other hand, a lot of interesting phenomena in time delays, including shape resonance \cite{heck2021attosecond} and two-center interference \cite{heck2022two} have been experimentally studied in this energy region, where the level of accuracy of the theory is partially limited by the reduced capability of the CC time delay models describing the relatively slow electrons. 
The general method for obtaining the $N$-photon transition amplitude from the $(N-1)$-photon transition amplitude has been known for a long time \cite{aymar1980two,shakeshaft1986sturmian}, which involves expressing the Green's function by the Sturmian expansion and extending into the continuum.
We shall show its application as a modification of expression (\ref{eq:M_two_photon}).
We give its analytical expression using the Appell's $F_1$ function, which quantitatively agrees with the TDSE and the experimental results in the whole energy range including the low-kinetic-energy region. We also note that the isotropic model can be extended to the $l$-dependent model by expanding the WKB approximation to the next term, and the Fano's propensity rule and the $l$-dependent CC transition phases can be well reproduced. 

This article is structured as following. In section \ref{sec:Methods}, we first define the functions for the intermediate and final states, then we derive the analytical formula for the transition amplitudes between the Tricomi's and Kummer's functions. In section \ref{sec:Results}, we show multiple results including the CC transition phases and time delays, and the Fano's propensity rule. These results are compared to TDSE and other theoretical approaches. 
We also compare our method to the experiments of linear-linear \cite{jiang2022atomic} and circular-circular \cite{han2024separation} RABBIT on helium.
We briefly discuss the behaviour of high $l$-states predicted by the analytical formula.
The conclusion and outlook can be found in section \ref{sec:Conclusion}.

\section{Method}
\label{sec:Methods}

\subsection{Regular and irregular solutions to the Coulomb wave equation}

The radial Coulomb electronic wave equation for the hydrogen atom (central charge $Z=1$) with asymptotic kinetic energy $E>0$ and angular momentum $l=0,1,2,\cdots$ reads (see \cite{DLMF}, section {(33.14)} and \cite{gaspard2018connection}, whereby equation (2) needs to be multiplied with $(-1)$ for describing an attractive potential): 
\begin{equation}
    \frac{\dd^2 w}{\dd r^2} + \left( 2E + \frac{2}{r} - \frac{l(l+1)}{r^2} \right) w = 0
    \label{eq:radial_equation}
\end{equation}
where $R(r) = w(r)/r$ is the radial wavefunction.
Equation (\ref{eq:radial_equation}) can be adapted for arbitrary $Z>0$ by transforming $\Tilde{E} = Z^2 {E}$ and $ \Tilde{r} = {r} / Z$, where $\Tilde{E}$ and $\Tilde{r}$ are the asymptotic kinetic energy and the radius, respectively. For simplicity, in the discussion below we let $Z=1$, while the treatment is general for any $Z>0$, and the corresponding result can be found in \ref{sec:multi_Z}. 
The regular solution of the outgoing radial wavefunction that satisfies the Coulomb wave equation can be written as (note that we have absorbed the factor $k^{l+1}$ into $C_l(k)$, compared to the definition in \cite{gaspard2018connection}):
\begin{eqnarray}
    R^{\mathrm{reg}}_{k,l}(r) \equiv f_l(k,r) = |C_l(k)| ~ r^l ~ 
    \exp \left( {\I k r} \right) 
    \Phi \left( l+1-\I / k, 2l+2, -2\I k r \right)
    \label{eq:f_l}
\end{eqnarray}
where $k=\sqrt{2E}$ is the momentum, and $\Phi$ is the confluent hypergeometric function (Kummer's function), also known as the ${_1F_1}$ function. Note that $f_l(k,r)$ is a real function under such definition. The normalisation factor for the regular solution of the Coulomb wave equation with the Coulomb phase included is expressed as:
\begin{equation}
    C_l(k) = {k}^{l+1} ~ {2}^l ~ \E^{{\pi}/{(2 k)}} ~ 
    \frac{\Gamma \left( l+1+\I / k \right)}{(2l+1)!}
    \label{eq:C_k_l}
\end{equation}
such that $r f_l(k,r)$ oscillates with amplitude of 1 at $r \rightarrow \infty$. 
Since equation (\ref{eq:radial_equation}) is a second-order differential equation, it has two linearly-independent solutions, and the other one can be expressed using the confluent hypergeometric function of the second kind $U(a,b;z)$, also known as the Tricomi's function: 
\begin{eqnarray}
    u_l(k,r) = B_l(k) ~ |C_l(k)| ~ r^l ~ 
    \exp \left( {\I k r} \right)
    U \left( l+1-\I / k, 2l+2, -2\I k r \right) ,
    \label{eq:u_l}
\end{eqnarray}
where we define the prefactor $B_l(k)$ as: 
\begin{equation}
    B_l(k) = -2 \I \E^{-{\pi}/{k}} {(-1)}^{l} \frac{(2l+1)!}{\Gamma\left( l+1+\I / k \right)}
\end{equation}
such that 
\begin{equation}
    f_l(k,r) = \Im \{ u_l(k,r) \} .
\end{equation}
Its real part is another real function that satisfies the Coulomb wave equation, also known as the irregular solution: 
\begin{equation}
    R^{\mathrm{irr}}_{k,l}(r) \equiv g_l(k,r) = \Re \{ u_l(k,r) \} .
\end{equation}
Note that $f_l(k,r)$ and $g_l(k,r)$ are linear combinations of $u_l(k,r)$ and its complex conjugate $u_l^*(k,r)$, which are two linearly independent solutions. 
Asymptotically, $u_l(k,r)$ approaches the spherical wave with a position-dependent phase (the Coulomb phase) \cite{gaspard2018connection}: 
\begin{eqnarray}
    \fl u_l(k,r \rightarrow \infty) = g_l(k,r \rightarrow \infty) + \I f_l(k,r \rightarrow \infty) \nonumber \\
    \approx {(-\I)}^{l}
    \frac{1}{r} \exp \left( \I k r + \frac{\I}{k} \ln(2kr) -\I \arg\{ C_l(k) \} \right) .
    \label{eq:u_l_asym}
\end{eqnarray}
Near the origin, the regular solution $f_l(k,r) \sim r^l$ approaches zero, while the irregular solution $g_l(k,r) \sim 1/r^{l+1}$ diverges. 
\begin{figure}[bt]
\begin{center}
\includegraphics[width=0.6\textwidth]{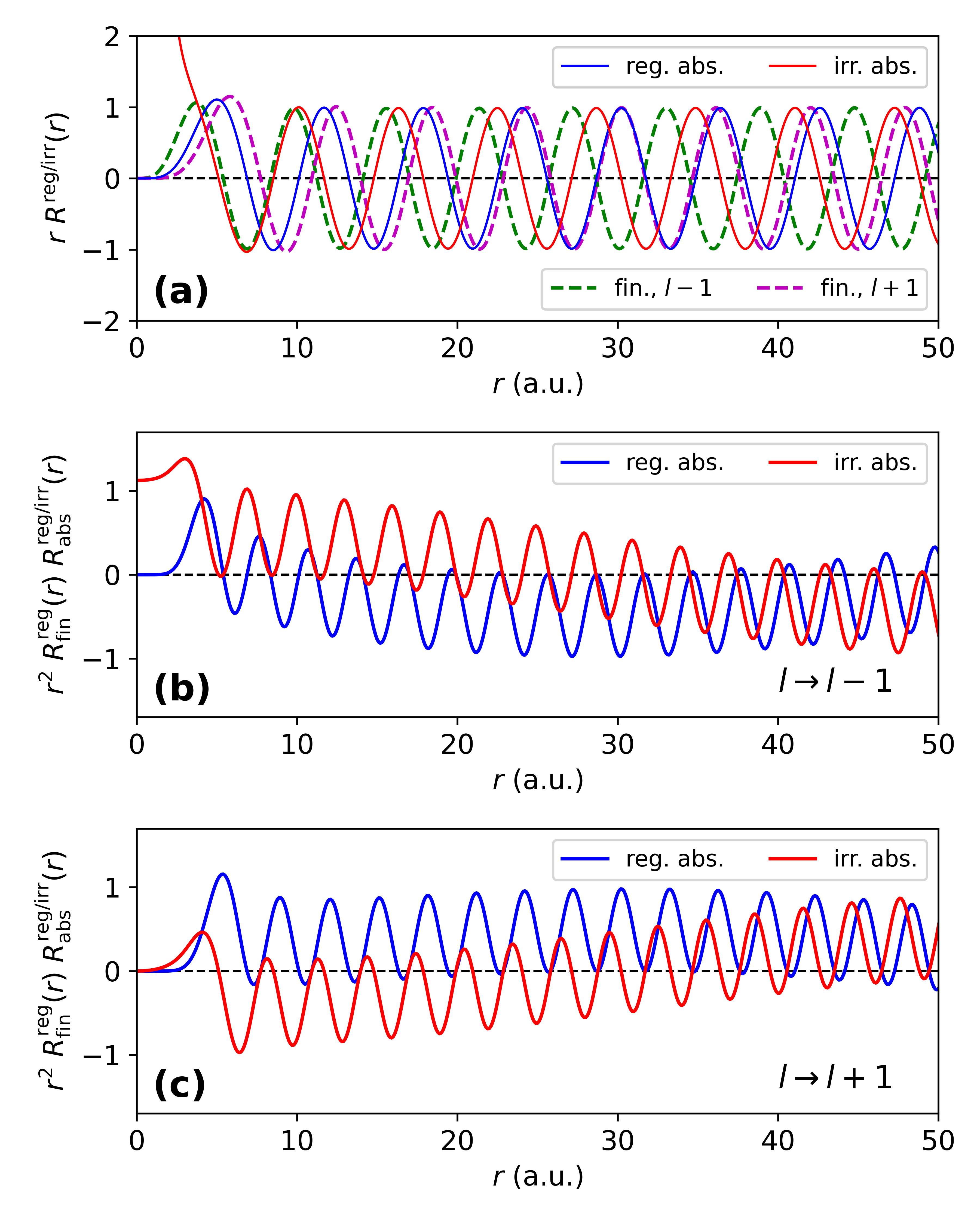}
\caption{(a) Radial wavefunctions (without the phase factors) of the final states (fin., dashed) at $E' = 15.00~\mathrm{eV}$ with $l' = 3$ (green) or $l' = 5$ (magenta), respectively, and  the intermediate state of the absorption pathway (abs., solid) at $E = 13.45~\mathrm{eV}$ with $l = 4$, where the photon energy of the dressing field is 1.55 eV. The regular (reg.) and irregular (irr.) parts of the intermediate state are plotted, while the final state has solely the regular part. (b) Products of the final-state radial wavefunction ($l'=3$) with the regular and the irregular parts of the intermediate-state radial wavefunction shown in (a), respectively. (c) The same as (b) but with the final state of $l'=5$. \label{fig:RadialWavefunction}}
\end{center}
\end{figure}
Although the irregular solution has infinitely large probability density near the origin, the product $r^2 f_l(k,r) g_{l \mp 1}(k,r) \sim r^{1 \pm 1}$ does not diverge near the origin, as plotted in figure \ref{fig:RadialWavefunction} (note that the normalisation for Coulomb waves here ensures that $r R(r)$ oscillates with modulus of 1 asymptotically), so the integral $\int_0^{R_{\mathrm{max}}>0} f_l(k,r) r g_{l \mp 1}(k,r) r^2 \dd r$ is finite. 
In the framework of the Green's function method, the two photon ionisation amplitude is proportional to the following integral \cite{shakeshaft1985note}: 
\begin{equation}
    \mathcal{I}^{+}_{l',l,\Tilde{l}}
    = \int_0^\infty \dd r \int_0^\infty \dd \Tilde{r}
    R_{k',l'}(r) r^2 \mathfrak{g}_l^+(r,\Tilde{r};E) {\Tilde{r}}^2 R_{\Tilde{n},\Tilde{l}}(\Tilde{r}) 
    \label{eq:I_two_photon_Green}
\end{equation}
where $R_{\Tilde{n},\Tilde{l}}(\Tilde{r})$ and $R_{k',l'}(r)$ are the radial wavefunctions of the initial (bound) and final (continuum) states, respectively, and $\mathfrak{g}_l^+(r,\Tilde{r};E)$ is the Green's function of the intermediate state that couples the initial and the final states. 
For the Coulomb potential, the initial and final states correspond to the regular solutions to the Coulomb equation. The radial Green's function in the three-dimensional space is \cite{meixner1933greensche,hostler1964coulomb,hostler1970coulomb}:
\begin{equation}
    \mathfrak{g}_l^+(r,\Tilde{r};E)
    = \frac{\Gamma(l+1-\I/k)}{2 \I k r_> r_<}
    W_{\I/k;l+1/2}(-2\I k r_>)
    \mathfrak{M}_{\I/k;l+1/2}(-2\I k r_<)
\end{equation}
where $k=\sqrt{2E}$ and $r_<$ ($r_>$) is the smaller (greater) variable between $r$ and $\Tilde{r}$. The Whittaker's functions $\mathfrak{M}_{\kappa;\mu}(z)$ and $W_{\kappa;\mu}(z)$ are related to the confluent hypergeometric functions according to \cite{DLMF}, section {(13.14)}
\begin{equation}
    \mathfrak{M}_{\kappa;\mu}(z)
    = \E^{-z/2} z^{\mu+1/2} \Phi(\mu + 1/2 - \kappa, 2 \mu + 1, z)
\end{equation}
and
\begin{equation}
    W_{\kappa;\mu}(z)
    = \E^{-z/2} z^{\mu+1/2} U(\mu + 1/2 - \kappa, 2 \mu + 1, z) .
\end{equation}
If we assume that the bound-continuum transition occurs mainly in the short range due to the finite spatial extension of the initial wavefunction, and the CC transition has major contribution from long range, we can replace $r_>$ and $r_<$ in the integral (\ref{eq:I_two_photon_Green}) by $r$ and $\Tilde{r}$, respectively. Thus, the two-photon transition amplitude is proportional to the product of
\begin{equation}
    \mathcal{I}_{l,\Tilde{l}}
    = \int_0^\infty f_l(k,\Tilde{r}) ~ \Tilde{r} ~ R_{\Tilde{n},\Tilde{l}}(\Tilde{r}) ~ {\Tilde{r}}^2 ~ \dd \Tilde{r}
\end{equation}
and 
\begin{equation}
    \mathcal{I}^{+}_{l',l}
    = \int_0^\infty f_{l'}(k',r) ~ r ~ u_l(k,r) ~ r^2 ~ \dd r
    \label{eq:I_f_r_u_Green}
\end{equation}
which corresponds to the bound-continuum and continuum-continuum transitions, respectively. Note that the additional term $r$ ($\Tilde{r}$) comes from the transform from the Whittaker's functions to the confluent hypergeometric functions.
Therefore, equation (\ref{eq:M_two_photon}) is slightly modified as: 
\begin{equation}
    \mathcal{M}_n(\vec{k}';\Omega)
    = \frac{1}{\I} 
    \lim_{\varepsilon \rightarrow 0^+}
    \sumint_{\nu}
    \frac{{\langle \vec{k}' | \hat{d}_{\omega} | \nu^+ \rangle}{\langle \nu | \hat{d}_{\Omega} | n \rangle}}{\epsilon_n + \Omega - \epsilon_{\nu} + \I \varepsilon}
    \label{eq:M_two_photon_mod}
\end{equation}
where for bound states $| \nu^+ \rangle = | \nu \rangle$ and for continuum states $| \nu^+ \rangle$ is the outgoing-wave solution corresponding to the standing-wave solution $| \nu \rangle$. The former relation is also indicated by the fact that the confluent hypergeometric functions of the first kind ($\Phi$) and second kind ($U$) truncate into the same associated Laguerre polynomials for the bound states. 
Compared to the approaches of Boll \textit{et al.} \cite{boll2022analytical} and Dahlstr{\"o}m \textit{et al.} \cite{dahlstrom12a}, integral (\ref{eq:I_f_r_u_Green}) uses the outgoing-wave solution to the Coulomb wave equation as the intermediate state instead of its asymptotic form.
Therefore, if the final state from one-photon ionisation has partial wave with angular momentum $l$ and asymptotically approaches $\frac{1}{r} \exp \left( \I k r \right)$, we choose the intermediate state as
\begin{eqnarray}
    \fl \Tilde{u}_l(k,r) = {\I}^l \E^{\I \arg\{ C_l(k) \}} u_l(k,r) 
    \nonumber \\
    =  {\I}^l B_l(k) ~ C_l(k) ~ r^l ~ 
    \exp \left( {\I k r} \right) 
    U \left( l+1-\I / k, 2l+2, -2\I k r \right)
\end{eqnarray}
and the final state as
\begin{eqnarray}
    \fl \I \Tilde{f}_{l'}(k',r) = {\I}^{l'} \E^{\I \arg\{ C_{l'}(k') \}} \I f_{l'}(k',r)
    \nonumber \\
    =  {\I}^{l'+1} ~ C_{l'}(k') ~ r^{l'} ~ 
    \exp \left( {\I k' r} \right)
    \Phi \left( {l'}+1-\I / k', 2{l'}+2, -2\I k' r \right)
\end{eqnarray}
where $(l,k)$ and $(l',k')$ correspond to the intermediate state and the final state, respectively. For the RABBIT experiment, where a photon from the dressing field is either absorbed (absorption pathway) or emitted (emission pathway), the final kinetic energy is $E' = k'^2 / 2 = E ~{\pm}^{\abs}_{\emi} ~ \omega$, while $ l' = l \pm 1$ can have both signs for each pathway, unless restricted by the magnetic quantum number in circular fields. For example, for the $1s$ electron in hydrogen, if a linearly polarised XUV photon ionises it to the $E p_0$ continuum state ($z$-axis defined as the polarisation direction), and the dressing field is co-linearly polarised, both absorption and emission pathways lead to the superposition of $E' s_0$ and $E' d_0$ states as the final state; on the other hand, if it is ionised by a circularly polarised XUV photon and has $E p_{+1}$ as the intermediate state ($z$-axis defined as the light propagation direction), and the dressing field is also circular with co-rotating polarisation, then for the absorption pathway only $E' d_{+2}$ is allowed, while the emission pathway leads to the superposition of the $E' s_0$ and $E' d_0$ states.
The dipole transition amplitude between the intermediate state and the final state is $\mathcal{E}_{\omega} \mathcal{T}_{l \rightarrow l'}(k,k')$, with
\begin{equation}
    \mathcal{T}_{l \rightarrow l'}(k,k') = -\pi N_k N_{k'} \int_0^\infty  
    \frac{1}{\I} \Tilde{f}_{l'}^*(k',r) ~ r ~ \Tilde{u}_{l}(k,r) ~ r^2 \dd r
    \label{eq:CC_T}
\end{equation}
where $N_{k} = \sqrt{2/(\pi k)}$ and $N_{k'} = \sqrt{2/(\pi k')}$ are the flux-based normalisation factors.
In another recent work \cite{benda2024angular}, Benda \textit{et al.} employ a similar Green's function method and reach an expression that is essentially the same as (\ref{eq:CC_T}), in the context of multiphoton ionisation including the RABBIT scheme with higher-order interference pathways. The challenge for the community so far is to convert this integral into a closed-form expression, or to numerically evaluate it as accurately as possible.
In the following sections, we show that equation (\ref{eq:CC_T}) and related integrals can indeed be expressed as analytical expressions. We will also give a simpler approximation that describes the asymptotic behaviour with consideration of the centrifugal energy following the WKB approximation.

\subsection{General expression of integral containing the product of the confluent hypergeometric functions of the first and the second kind}

Using the definition of $\Tilde{f}_{l'}(k',r)$ and $\Tilde{u}_{l}(k,r)$, we can write equation (\ref{eq:CC_T}) into: 
\begin{eqnarray}
    \fl \mathcal{T}_{l \rightarrow l'}(k,k')  =  -\pi N_k N_{k'} \I^{l-l'-1} B_l(k) C_l(k) C_{l'}^*(k') \nonumber \\
    \times \lim_{Q_0 \rightarrow 0^+} \int_0^\infty
    \exp \left[ \left( {\I k} + {\I k'} - Q_0 \right) r \right] r^{l+l'+3}  \nonumber \\
    \times U \left( l+1-\I / k, 2l+2, -2\I k r \right) \nonumber \\
    \times \Phi \left( {l'}+1-\I / k', 2{l'}+2, -2\I k' r \right) \dd r
    \label{eq:T_CC_expanded}
\end{eqnarray}
where $Q_0$ is a small positive number that ensures the integral converges according to the Abel-Dirichlet test \cite{Zorich2016}. Note that we have used the fact that $\exp \left( {\I k r} \right) \times \Phi \left( l+1-\I / k, 2l+2, -2\I k r \right)$ is real and equal to its complex conjugate.
Although the similar integral involving two Kummer's functions has already been discussed by Gordon almost a century ago \cite{gordon1929berechnung} and the more general case is known to have the analytical formulation using the Appell's functions \cite{tarasov1995multipole,tarasov2003w,saad2003integrals}, to the best of our knowledge, the formula for the integral involving the product of the confluent hypergeometric functions of the first and second kind has yet to be explicitly demonstrated. Here we derive the analytical expression following the spirit of Gordon \cite{gordon1929berechnung}.
Let us consider the integral with a more general form: 
\begin{eqnarray}
    \mathcal{J}^{\sigma}_{\rho} (a,b,\lambda;n',\lambda';Q)
    = \int_0^\infty \E^{-Q\xi} \xi^{\rho+\sigma} 
    U(a,b;\lambda \xi) \Phi(-n',\rho+1;\lambda' \xi) \dd \xi .
\end{eqnarray}
The confluent hypergeometric function of the second kind has the following recurrence relations \cite{DLMF}, equation {(13.3.10)}:
\begin{eqnarray}
    U(a,b;z) = && \frac{1}{z}\big[ (b-a-1)U(a,b-1;z) + U(a-1,b-1;z) \big] .
    \label{eq:U_recurrence}
\end{eqnarray}
Hence, we have the relation: 
\begin{eqnarray}
    \fl \mathcal{J}^{\sigma}_{\rho} (a,b,\lambda;n',\lambda';Q) 
    = \frac{1}{\lambda} \big[
    (b-a-1)\mathcal{J}^{\sigma-1}_{\rho} (a,b-1,\lambda;n',\lambda';Q) 
    \nonumber \\
    + \mathcal{J}^{\sigma-1}_{\rho} (a-1,b-1,\lambda;n',\lambda';Q) \big] .
    \label{eq:J_sigma}
\end{eqnarray}
For $\sigma > 0$, which is the case for the transition between two continuum states (from equation (\ref{eq:T_CC_expanded}) we have $l-l'+2 = \pm 1 + 2$), the integral finally comes to $\sigma=0$ terms: 
\begin{eqnarray}
    \mathcal{J}^{0}_{\rho} (a,b,\lambda;n',\lambda';Q) 
    = \int_0^\infty \E^{-Q\xi} \xi^{\rho} 
    U(a,b;\lambda \xi) \Phi(-n',\rho+1;\lambda' \xi) \dd \xi.
\end{eqnarray}
In particular, we can write the two expressions for $\sigma = $ 1 and 3 explicitly: 
\begin{eqnarray}
    \fl \mathcal{J}^{1}_{\rho} (a,b,\lambda;n',\lambda';Q) 
    = \frac{1}{\lambda}\big[
    (b-a-1)\mathcal{J}^{0}_{\rho} (a,b-1,\lambda;n',\lambda';Q) \nonumber \\
    + \mathcal{J}^{0}_{\rho} (a-1,b-1,\lambda;n',\lambda';Q) \big]
    \label{eq:J_1}
\end{eqnarray}
and
\begin{eqnarray}
    \fl \mathcal{J}^{3}_{\rho} (a,b,\lambda;n',\lambda';Q)
    = \frac{1}{\lambda^3}\big[
    (b-a-3)(b-a-2)(b-a-1) \mathcal{J}^{0}_{\rho} (a,b-3,\lambda;n',\lambda';Q) \nonumber \\
    + 3 (b-a-2)(b-a-1) \mathcal{J}^{0}_{\rho} (a-1,b-3,\lambda;n',\lambda';Q) \nonumber \\
    + 3 (b-a-1) \mathcal{J}^{0}_{\rho} (a-2,b-3,\lambda;n',\lambda';Q) \nonumber \\
    + \mathcal{J}^{0}_{\rho} (a-3,b-3,\lambda;n',\lambda';Q)
    \big] .
    \label{eq:J_3}
\end{eqnarray}
The base case $\mathcal{J}^{0}_{\rho} (a,b,\lambda;n',\lambda';Q)$ can be expressed by the Appell's $F_1$ function \cite{functionswolfram}, and the detailed derivation can be found in \ref{sec:Appendix_integral}:
\begin{eqnarray}
\fl \mathcal{J}^{0}_{\rho} (a,b,\lambda;n',\lambda';Q) = 
\frac{\Gamma(\rho-b+2)}{\Gamma(\rho-b+a+2)}
\frac{\rho! ~ {(Q-\lambda')}^{n'}}{Q^{\rho+1+n'}} \nonumber \\
\times
F_1 \left( a;\rho+1+n',-n';\rho-b+a+2;1-\frac{\lambda}{Q},1-\frac{\lambda}{Q-\lambda'} \right) \nonumber \\
= \frac{\Gamma(\rho-b+2)}{\Gamma(\rho-b+a+2)}
\frac{\rho!}{\lambda^{\rho+1}} \nonumber \\
\times
F_1 \left( \rho-b+2; \rho+1+n', -n'; \rho-b+a+2; 1 - \frac{Q}{\lambda}, 1 - \frac{Q-\lambda'}{\lambda} \right)
\label{eq:J_0_rho_Q_in_F1}
\end{eqnarray}
where in the last step we used the relation \cite{Bateman:100233}, equation {(5.11.1)}: 
\begin{eqnarray}
    \fl F_1(\alpha;\beta_1,\beta_2;\gamma;z_1,z_2) 
    = {(1-z_1)}^{-\beta_1} {(1-z_2)}^{-\beta_2} 
    F_1 \left( \gamma-\alpha;\beta_1,\beta_2;\gamma;\frac{z_1}{z_1-1},\frac{z_2}{z_2-1} \right) .
    \label{eq:appellf1_transform}
\end{eqnarray}
With help of equations (\ref{eq:J_sigma}, \ref{eq:J_1}, \ref{eq:J_3}), we can calculate the multipole transition amplitudes between states expressed by the confluent hypergeometric functions of the first and the second kind.
Therefore, equation (\ref{eq:T_CC_expanded}) can be expressed as: 
\footnote{Here we make a special note that although $Q_0$ in equation (\ref{eq:T_CC_in_F1}) can be exactly zero, where the last two arguments in equation (\ref{eq:J_0_rho_Q_in_F1}) are real and can be evaluated by analytic continuation when one of them has modulus greater than 1. However, due to the ambiguous phase of $(-1)$ that occurs upon analytic continuation, one of the branches of the parameters may become incorrect, which means that taking the complex conjugates of all the first four arguments in $F_1$ does not return the complex conjugate of the result, although from its definition one would expect so. Therefore, it is recommended to numerically confirm that the case of $Q_0 = 0$ is indeed the limit of $Q_0 = 0^+$.}
\begin{eqnarray}
\label{eq:T_CC_in_F1}
\fl \mathcal{T}_{l \rightarrow l'}(k,k') = -\pi N_k N_{k'} \I^{l-l'-1} B_l(k) C_l(k) C_{l'}^*(k') \\
\times \lim_{Q_0 \rightarrow 0^+} \mathcal{J}^{l-l'+2}_{2l'+1} \left( l+1-\frac{\I}{k},2l+2,-2\I k;-\left( l'+1-\frac{\I}{k'} \right),-2\I k'; Q_0 -\I(k+k') \right) \nonumber .
\end{eqnarray} 
This equation provides an analytical expression for the radial integral of the CC transition. 
One can find the similarity between our equations (\ref{eq:J_sigma}, \ref{eq:J_1}, \ref{eq:J_3}, \ref{eq:J_0_rho_Q_in_F1}) and the equation (3.1) in \cite{saad2003integrals}. In fact, we can write 
\begin{eqnarray}
    \fl B_l(k) |C_l(k) C_{l'}(k')| 
    \mathcal{J}^{l-l'+2}_{2l'+1} \left( l+1-\frac{\I}{k},2l+2,-2\I k;  -\left( l'+1-\frac{\I}{k'} \right),-2\I k';Q_0 -\I(k+k') \right)
    \nonumber \\
    = \int_0^\infty f_{l'}(k',r) ~ r \E^{-Q_0 r} ~ u_l(k,r) ~ r^2 \dd r 
    \\
    = \int_0^\infty f_{l'}(k',r) ~ r \E^{-Q_0 r} ~ g_l(k,r) ~ r^2 \dd r
    + \I \int_0^\infty f_{l'}(k',r) ~ r \E^{-Q_0 r} ~ f_l(k,r) ~ r^2 \dd r \nonumber
    \label{eq:J_reg_irr}
\end{eqnarray}
and one immediately sees that the imaginary part of equation (\ref{eq:J_reg_irr}) is the transition amplitude between two regular solutions to the Coulomb wave equation, also known as the Gordon's integral \cite{tarasov1995multipole,tarasov2003w,saad2003integrals}.
Note that our formula has similarity to the general expression for the double- or multi-photon ionisation amplitude from bound states of hydrogen atoms given in \cite{karule1978two,karule2003general}, which use the Sturmaian expansion with modifications in the continuum.

\subsection{Asymptotic formula for the continuum-continuum transition amplitude with the inclusion of centrifugal potential}

The purpose of this section is to demonstrate the mechanism with simpler mathematical expressions. Such approaches are usually available for describing the asymptotic behaviour at high kinetic energies, while we try to make the compromise that the formula applied at low kinetic energies should at least give a semi-quantitative description, such that the main features are still captured. Let us consider the phase contribution of the centrifugal potential under the WKB approximation. The phase factor of the outgoing wave can be written as: 
\begin{eqnarray}
    \fl  S_l(k,r;Z) = \int_0^r \sqrt{\frac{2Z}{\mathfrak{r}} - \frac{2 b}{\mathfrak{r}^2} + k^2} ~ \dd \mathfrak{r} \nonumber \\
    = \int_0^r k + \frac{\mathfrak{r}^{-1} Z}{k} - \frac{\mathfrak{r}^{-2} (2k^2 b + Z^2)}{2 k^3} + \mathcal{O}(\mathfrak{r}^{-3}) ~ \dd \mathfrak{r} \nonumber \\
    = k r + \ln(2kr) Z/ k + r^{-1} \frac{2k^2 b + Z^2}{2 k^3} + \mathcal{O}(r^{-2})
\end{eqnarray}
where $b = l(l+1)/2$ comes from the centrifugal potential, and the expansion is made at $r \rightarrow \infty$. Thus, the outgoing radial wave function can be estimated as: 
\begin{eqnarray}
    \fl r R_{k,l}(r) \approx \exp(\I S_l(k,r;Z)) \nonumber \\
    = \E^{\I k r} {(2kr)}^{\I Z / k} \exp(\I q r^{-1} + \mathcal{O}(r^{-2})) \nonumber \\
    = \E^{\I k r} {(2kr)}^{\I Z / k} \left( 1 + \I q r^{-1} + \mathcal{O}(r^{-2}) \right)
\end{eqnarray}
where $q = (2k^2 b + Z^2) / (2k^3)$ scales with $k^{-3}$ for the $s$ ($l=0$) wave, but with $k^{-1}$ for waves with higher angular momenta. This explains why the phase contribution of the centrifugal potential should not be neglected, as $k^{-1}$ is a rather slow convergence under the experimental context (typically $E = {10}^{1} \sim {10}^{2} ~ \mathrm{eV}$, $k = 1 \sim 3 ~ \mathrm{a.u.}$). The dipole transition between two outgoing waves can thus be approximated as: 

\begin{eqnarray}
    \fl \mathcal{T}^{(\mathrm{P})}_{l \rightarrow l'}(k,k';r_0;Z) \approx
    - \frac{\pi}{2} N_k N_{k'} \frac{{(2k)}^{\I Z/k}}{{(2k')}^{\I Z/k'}} 
    \int_{r_0}^\infty \E^{-\I (k' - k) r} ~ r^{\I Z/k - \I Z/k'} \left[ r + \I(q - q') + \mathcal{O}(r^{-1}) \right] ~ \dd r
    \nonumber \\
    \approx -\frac{\pi}{2} N_k N_{k'} \frac{{(2k)}^{\I Z/k}}{{(2k')}^{\I Z/k'}} 
    \left[ \frac {\Gamma(s,\Lambda r_0)}{\Lambda^{s}} + \I (q - q') \frac {\Gamma(s-1,\Lambda r_0)}{\Lambda^{s-1}} \right]
    \label{eq:T_CC_in_gamma}
\end{eqnarray}
where $s = 2 + \I Z/k - \I Z/k'$ and $\Lambda = \I (k' - k)$. 
The factor of $1/2$ comes from the fact that the final state represented by the Kummer's function contains the $k'$- and $(-k')$-parts, while the $(-k')$-part practically vanishes due to the large imaginary part in the gamma function \cite{dubuc1990approximation}.
Here we have used the incomplete gamma function 
\begin{equation}
    \Gamma(s,x) = \int_x^\infty t^{s-1} \E^{-t} ~ \dd t
\end{equation}
to include the case where the integral starts from $r_0 > 0$ instead of 0. This can be useful for the more complex scattering problem, where the (time-dependent) Schr{\"o}dinger equation is numerically solved inside the radius $r_0$, while the outgoing electron wavefunction is extrapolated beyond $r_0$ using the $l$-dependent asymptotic formula, assuming that the potential is converged except the Coulomb ($\sim \mathcal{O}(r^{-1})$) and centrifugal-like ($\sim \mathcal{O}(r^{-2})$) parts. 
When $r_0 = 0$, $\Gamma(s,0) = \Gamma(s)$ is the gamma function, and $\Gamma(s) / \Gamma(s-1) = s-1$. 
The first term in equation (\ref{eq:T_CC_in_gamma}) corresponds to equation (\ref{eq:T_CC_iso_P}) reported by Dahlstr{\"o}m \textit{et al.} \cite{dahlstrom13a,dahlstrom12a,dahlstrom12b}, and the second term depends on the angular momenta of the initial and the final states. 

Note that Dahlstr{\"o}m \textit{et al.} also reported another formula (equation (\ref{eq:T_CC_iso_PA})) including the long-range amplitude correction according to the WKB approximation \cite{dahlstrom12a}: 
\begin{equation}
    r R_{k,l}(r) \approx \exp(\I S_l(k,r;Z))
    {k}^{1/2} {\left( 2Z/r + k^2 \right)}^{-1/4}
\end{equation}
where the amplitude factor is
\begin{equation}
    {\left( \frac{2Z}{k^2 r} + 1 \right)}^{-1/4}
    = 1 - \frac{Z}{2 k^2 r} + \mathcal{O}(r^{-2}) .
    \label{eq:WKB_amp_factor}
\end{equation}
Following the similar treatment, equation (\ref{eq:T_CC_in_gamma}) is modified as 
\begin{eqnarray}
    \fl \mathcal{T}^{(\mathrm{P+A})}_{l \rightarrow l'}(k,k';r_0;Z) \nonumber \\
    \approx
    - \frac{\pi}{2} N_k N_{k'} \frac{{(2k)}^{\I Z/k}}{{(2k')}^{\I Z/k'}} 
    \left[ \frac {\Gamma(s,\Lambda r_0)}{\Lambda^{s}} + ( \I q - \I q' - Q - Q' ) \frac {\Gamma(s-1,\Lambda r_0)}{\Lambda^{s-1}} \right]
    \label{eq:T_CC_in_gamma_PA}
\end{eqnarray}
where $q = (2k^2 b + Z^2) / (2k^3) = (l (l+1) k^2 + Z^2) / (2k^3)$ and $Q = Z/(2 k^2)$. One can see the parallelity between these terms and the parameter defined in equations (13) and (16a) of reference \cite{berkane2024probing}, which is responsible for the $l$-dependence of the radial integral. The benefit of the 
modified asymptotic approximation proposed in this work is that it allows $r_0 > 0$, which can be used for extending the numerical calculation from a finite cell into infinity.
The superscripts $\mathrm{(P)}$ and $\mathrm{(P+A)}$ in equations (\ref{eq:T_CC_in_gamma}, \ref{eq:T_CC_in_gamma_PA}) refer to including the phase variation and the phase and amplitude variations, respectively. 
For the hydrogen atom at low kinetic energies, the phase computed by TDSE typically lies between equation (\ref{eq:T_CC_in_gamma}) and equation (\ref{eq:T_CC_in_gamma_PA}) with $r_0 = 0$. Although the agreement may improve by choosing a suitable $r_0$ in a similar fashion as in the (P+A$^\prime$) model of reference \cite{dahlstrom12a}, it is better to evaluate according to equation (\ref{eq:T_CC_in_F1}).

\subsection{Numerical evaluation}

For numerical evaluation in this work, we used \texttt{Python 3.8.10} \cite{oliphant2007python} with \texttt{numpy 1.20.2} \cite{harris2020array} and \texttt{mpmath 1.1.0} \cite{mpmath}. The latter package is capable of calculating many special functions with complex arguments.

\section{Results and discussion}
\label{sec:Results}

\subsection{Continuum-continuum transition phase for the hydrogen atom}
\label{sec:CC_phase}

\begin{figure}[bt]
\begin{center}
\includegraphics[width=1\textwidth]{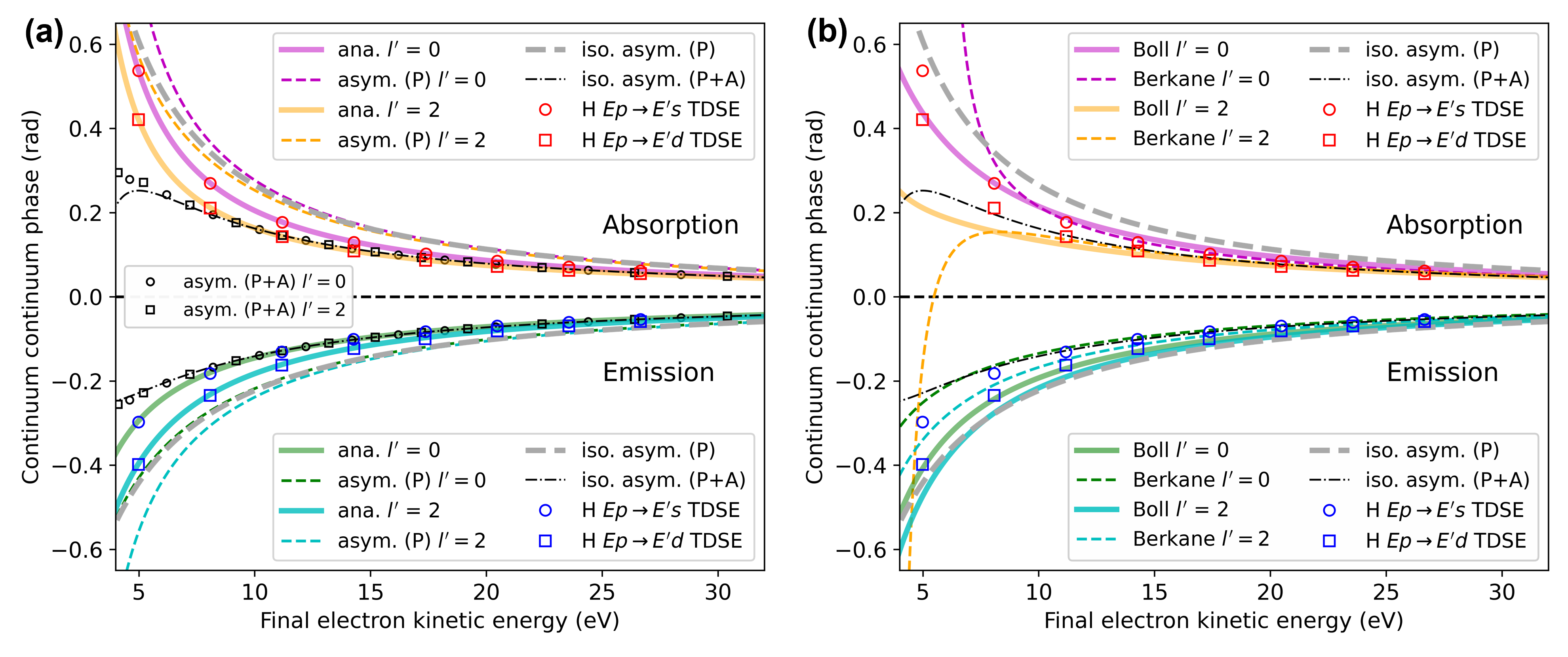}
\caption{The phase of the CC transition (with dressing field wavelength of 800 nm) for the hydrogen atoms from the intermediate state of $l=1$ to the final state of $l'= 1 \mp 1$. The TDSE results are taken from \cite{boll2022analytical}, which practically overlap with the SOPT calculation; the isotropic asymptotic approximation with phase correction (iso. asym. (P)) and with phase and amplitude correction (iso. asym. (P+A)) are taken from \cite{dahlstrom12a}. (a) The analytical formula (ana.) corresponds to (\ref{eq:T_CC_in_F1}); the $l$-dependent asymptotic approximation with phase correction (asym. (P), $r_0=0$) corresponds to (\ref{eq:T_CC_in_gamma}); the $l$-dependent asymptotic approximation with phase and amplitude correction (asym. (P+A), $r_0=0$) corresponds to (\ref{eq:T_CC_in_gamma_PA}). (b) The formulae proposed by Boll \textit{et al.} \cite{boll2022analytical} and by Berkane \textit{et al.} \cite{berkane2024probing} are labelled as ``Boll'' and ``Berkane'', respectively. \label{fig:TransitionPhase} }
\end{center}
\end{figure}

In order to test the accuracy of the analytical formula and its asymptotic approximations, we compare to the time-dependent Schr{\"o}dinger equation (TDSE) results reported in \cite{dahlstrom12a,boll2022analytical}, which are currently considered to be the benchmark for the CC transition amplitudes. 
As shown in figure \ref{fig:TransitionPhase} (a), our equation (\ref{eq:T_CC_in_F1}) perfectly agrees with the TDSE values, which manifests the validity of our approach. Regarding the asymptotic approximations, the (P) and (P+A) formulae given by Dahlstr{\"o}m \textit{et al.} are independent of the angular momentum $l'$ (isotropic), and the CC phases of $p \rightarrow s$ and $p \rightarrow d$ lie between the two approximations. The modified asymptotic approximation (P) given by equation (\ref{eq:T_CC_in_gamma}) reproduces the phase difference between the $p \rightarrow s$ and $p \rightarrow d$ channels, indicating that the effect of the centrifugal potential is indeed captured by the expansion, although they both overestimate the absolute value of the phase at low kinetic energies, as in the case of the isotropic formula. On the other hand, the modified asymptotic approximation (P+A) given by equation (\ref{eq:T_CC_in_gamma_PA}) show very little difference between $p \rightarrow s$ and $p \rightarrow d$, and both of them are well aligned with the corresponding isotropic formula. This is because at low kinetic energies the amplitude-variation effect overrides the phase-variation effect, while the former is independent of $l'$ under the far-field expansion up to $\mathcal{O}(r^{-1})$. In figure \ref{fig:TransitionPhase} (b), we compare two alternative formulae reported by Boll \textit{et al.} \cite{boll2022analytical}, equation (10) and by Berkane \textit{et al.} \cite{berkane2024probing}, equation (16), respectively. They both reproduce the correct relative relation between $(l \rightarrow l+1)$ and $(l \rightarrow l-1)$, while in general the formula proposed in this work agrees with TDSE best.

\subsection{Fano's propensity rule in continuum-continuum transition}
\label{sec:CC_Fano}

\begin{figure}[bt]
\begin{center}
\includegraphics[width=1\textwidth]{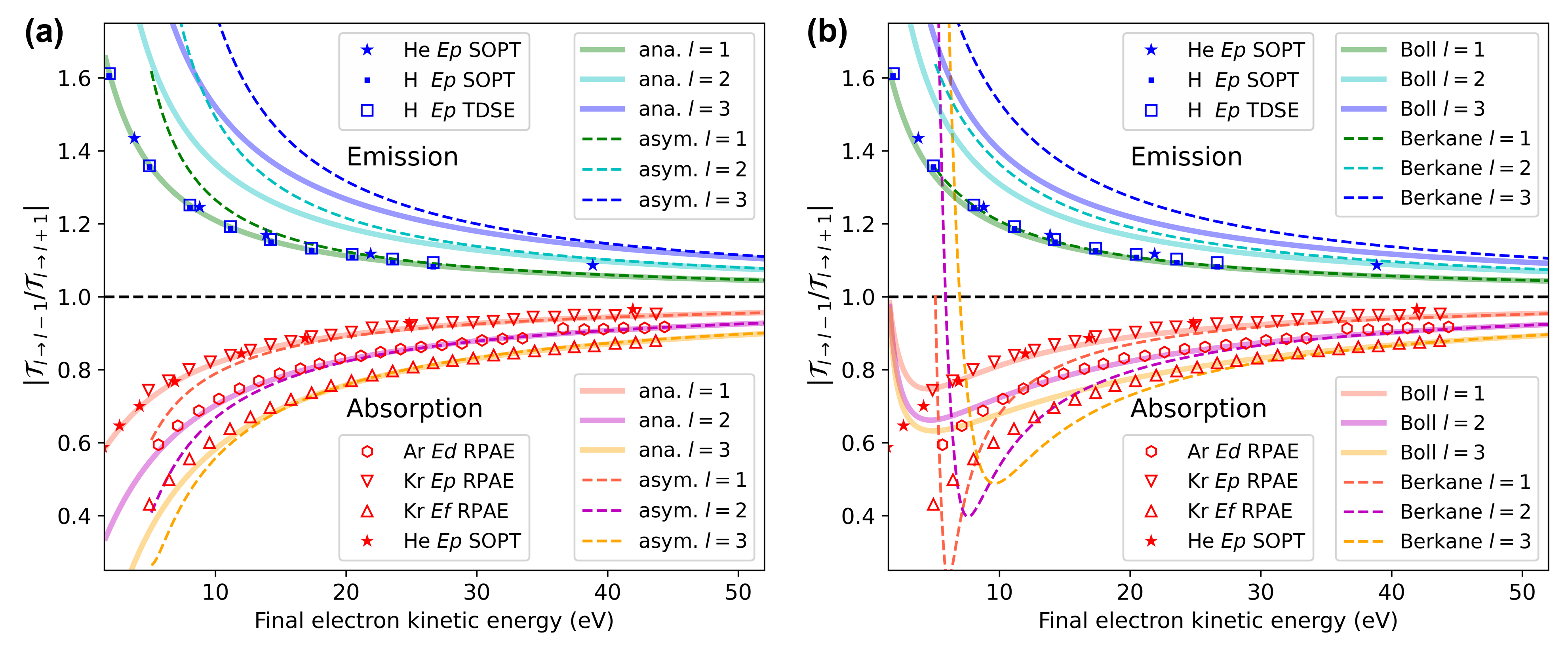}
\caption{Moduli ratios for the absorption and emission pathways (with dressing field wavelength of 800 nm). The computational results (RPAE, SOPT, TDSE) are taken from \cite{boll2022analytical}, where the RPAE results originate from \cite{busto2019fano}. The angular momentum $l$ corresponds to the intermediate state. (a) The analytical formula (ana.) corresponds to (\ref{eq:T_CC_in_F1}) and the results are shown by the thick solid curves, while the results from the $l$-dependent asymptotic approximation (\ref{eq:T_CC_in_gamma}, \ref{eq:T_CC_in_gamma_PA}) are shown by the thin dashed curves with the corresponding colours, where the results from the (P) and (P+A) formulae almost overlap, so only the ones from (P) are shown. (b) The formulae proposed by Boll \textit{et al.} \cite{boll2022analytical} and by Berkane \textit{et al.} \cite{berkane2024probing} are labelled as ``Boll'' and ``Berkane'', respectively. \label{fig:TransitionAbsRatio}}
\end{center}
\end{figure}

Fano's propensity rule in CC transition reported in references \cite{busto2019fano,boll2022analytical} is tested using our equation (\ref{eq:T_CC_in_F1}) and its asymptotic approximations, as shown in figure \ref{fig:TransitionAbsRatio} (a). 
The Fano's propensity rule would be entirely absent if one used the isotropic asymptotic formulae, since they do not have $l$-dependence.
The analytical formula proposed in this work again shows excellent agreement with TDSE \cite{dahlstrom12a,boll2022analytical}, SOPT \cite{jayadevan2001two}, and RPAE \cite{busto2019fano} for intermediate states with various angular momenta. Note that the modified asymptotic approximation shows very good agreement from about 15 eV, where the absorption pathway converges seemingly faster than the emission pathway. The (P) and (P+A) approximations ($r_0=0$) give very similar moduli ratios, as the amplitude variation effect is mostly cancelled in the ratios. 
In figure \ref{fig:TransitionAbsRatio} (b) we compare the numerical values with the analytical formulae proposed by Boll \textit{et al.} \cite{boll2022analytical} and Berkane \textit{et al.}, respectively. In general, at higher kinetic energies ($\gtrsim 30~\mathrm{eV}$), all of the four $l$-dependent formulae agree with the numerical computation very well. However, at lower kinetic energies, particularly below 10 eV, Boll \textit{et al.}'s and Berkane \textit{et al.}'s formulae show clear deviation from the numerical results for the absorption pathway, and for the asymptotic approximation reported in this work the deviation occurs for both absorption and emission pathways. On the other hand, the analytical formula (\ref{eq:T_CC_in_F1}) in this work remains good agreement below 10 eV. This indicates that for analytical approaches that utilises the asymptotic behaviour of the radial wave function instead of using the exact Coulomb wave function, the result is only reliable for $k \gtrsim 1~\mathrm{a.u.}$.

\subsection{Continuum-continuum transition time delay}
\label{sec:CC_time_delay}

\begin{figure}[bt]
\begin{center}
\includegraphics[width=0.9\textwidth]{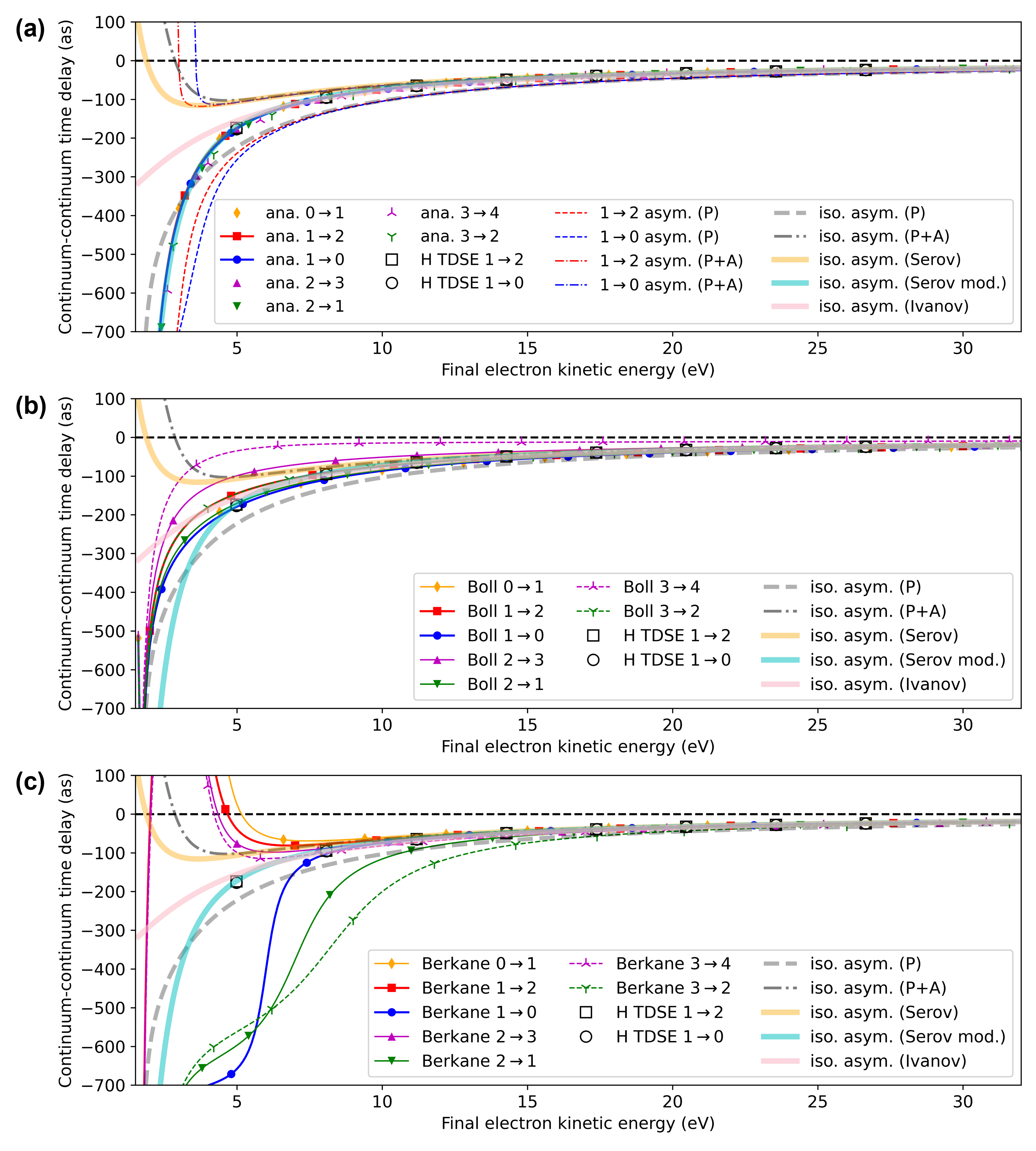}
\caption{The CC time delays obtained from the phases shown in figure \ref{fig:TransitionPhase}. The TDSE results from hydrogen atoms are taken from \cite{boll2022analytical}; the isotropic asymptotic approximation are taken from \cite{dahlstrom12a} (``P'' and ``P+A''), \cite{serov2015interpretation} (``Serov'' and ``Serov mod.''). (a) The markers and curves ``ana.'' are given by the analytical formula (\ref{eq:T_CC_in_F1}) labelled as ``(intermediate angular momentum) $\rightarrow$ (final angular momentum)''. (b) The markers and curves ``Boll'' are given by the $l$-dependent formula in \cite{boll2022analytical}. (c) The markers and curves ``Berkane'' are given by the $l$-dependent formula in \cite{berkane2024probing}. \label{fig:TransitonDelayCombined}}
\end{center}
\end{figure}

The CC transition time delay defined by the finite difference is
\begin{equation}
    \tau_{\CC}(E) = \frac{1}{2 E_{\CC}} \left( \phi^{\emi}(E_+ \rightarrow E) - \phi^{\abs}(E_- \rightarrow E) \right)
    \label{eq:T_CC_finite_diff_def}
\end{equation}
where $E_\pm = E \pm E_{\CC}$ is the electron kinetic energy of the intermediate state, and $\phi$ is the phase of the transition amplitude. 
The results given by the analytical formula (\ref{eq:T_CC_in_F1}) are plotted in figure \ref{fig:TransitonDelayCombined} (a). The values for the $E p \rightarrow E' s$ and $E p \rightarrow E' d$ channels agree perfectly with the TDSE for hydrogen atoms. This again manifests the validity of our analytical formula. The (P) and (P+A) formulae reported by Dahlström \textit{et al.} \cite{dahlstrom12a} and with the modification introduced in this work lie below and above the curves, respectively. In particular, the (P+A) formula bends to the opposite direction at lower kinetic energies, although it matches the TDSE better at higher kinetic energies. 
Interestingly, although the phases of different $l$-channels vary, as shown in figure \ref{fig:TransitionPhase}, the phase differences between the emission and absorption pathways ($l \leq 3$) are very close, as pointed out by Busto \textit{et al.} \cite{Busto2024Atomic}. Hence, the key problem is to correctly determine the relative phase offset between channels and their moduli ratios, which can be accurately achieved by the analytical formula, as demonstrated in the previous sections. 
In figure \ref{fig:TransitonDelayCombined} (b) and (c), we compare the CC transition time delays given by Boll \textit{et al.}'s \cite{boll2022analytical} and Berkane \textit{et al.}'s \cite{berkane2024probing} $l$-dependent formulae, respectively. The closeness among different $l$-channels reported in \cite{Busto2024Atomic} is no longer observed. This manifests that the analytical formula reported in this work further improves the accuracy of the CC transition time delays compared to the previously reported ones.

We further compare several alternative formulae that have been derived for the Coulomb-laser coupling (CLC) time delay, i.e. equation (14.44) from \cite{serov2015interpretation} (``Serov'') and equation (13) from \cite{ivanov2011accurate} (``Ivanov''). They do not have explicit $l$-dependence and are all labelled as ``iso. asym.'' in figure \ref{fig:TransitonDelayCombined}. 
Ivanov \textit{et al.}'s formula (for $Z=1$) reads:
\begin{equation}
    \tau_{\rm CLC}^{\rm (Ivanov)} = 
    -\frac{1}{k^3 + \omega \pi / 2}
    \left[ \ln \left( \frac{a k^2}{\omega} \right) - \gamma_{\rm Euler} + \frac{\pi}{2} \frac{\omega}{a k^2} \right]
    \label{eq:tau_CLC_Ivanov}
\end{equation}
where $k$ is the electron's final momentum, $\omega = E_{\CC}$ is the dressing field frequency, and $\gamma_{\rm Euler}$ is the Euler's constant ($\approx 0.577$).
The factor $a$ in (\ref{eq:tau_CLC_Ivanov}) is defined as
\begin{equation}
    a = 2 \exp(-2 k \times \arg\{ \Gamma(1 - \I/k) \}) ~ .
\end{equation}
Serov \textit{et al.}'s formula reads: 
\begin{equation}
    \tau_{\rm CLC}^{\rm (Serov)} = 
    -\frac{Z}{k^3} \left[ \ln \left( \frac{2 k^2}{\omega} \right) - 1 - \gamma_{\rm Euler} \right]
    \label{eq:tau_CLC_Serov}
\end{equation}
Note that equation (\ref{eq:tau_CLC_Serov}) is equation (25) in reference \cite{berkane2024probing}, which is derived as the soft-photon limit of the CC transition delay with $\omega \rightarrow 0$ under the (P+A) approximation in \cite{dahlstrom12a}. This indicates the fundamental consistency between CC and CLC time delays \cite{serov2015interpretation, kheifets2022ionization}. 
Besides, we have made an \textit{ad hoc} modification to Serov \textit{et al.}'s formula into:
\begin{equation}
    \tau_{\rm CLC}^{\rm (Serov, mod.)} = 
    -\frac{Z}{k^3} \left[ \ln \left( \frac{2 k^2}{\omega} \right) - 1 - \gamma_{\rm Euler} + \frac{3 \pi}{4} \frac{\omega Z}{k^3} \right]
    \label{eq:tau_CLC_Serov_mod}
\end{equation}
The modified formula (\ref{eq:tau_CLC_Serov_mod}) can be viewed as equation (14.43) in \cite{serov2015interpretation} with $\alpha = 0$, which decouples the CLC time delay from the Wigner time delay, while in the original formula (14.43) in \cite{serov2015interpretation} the CLC time delay has explicit dependency of the Wigner time delay. For kinetic energy $\geq 5~{\mathrm{eV}}$, both (\ref{eq:tau_CLC_Ivanov}) and (\ref{eq:tau_CLC_Serov_mod}) show good agreement with TDSE. In particular, formula (\ref{eq:tau_CLC_Serov_mod}) quantitatively agrees with the CC time delays (different $l$-channels yield similar CC time delays) given by the analytical formula (\ref{eq:T_CC_in_F1}) in the whole energy range, indicating that the last term in (\ref{eq:tau_CLC_Serov_mod}) functions as a correction to the soft-photon limit regarding the CC time delay.

\subsection{Angle-dependent RABBIT time delay for helium}
\label{sec:helium_circir}

\begin{figure}[bt]
\begin{center}
\includegraphics[width=1\textwidth]{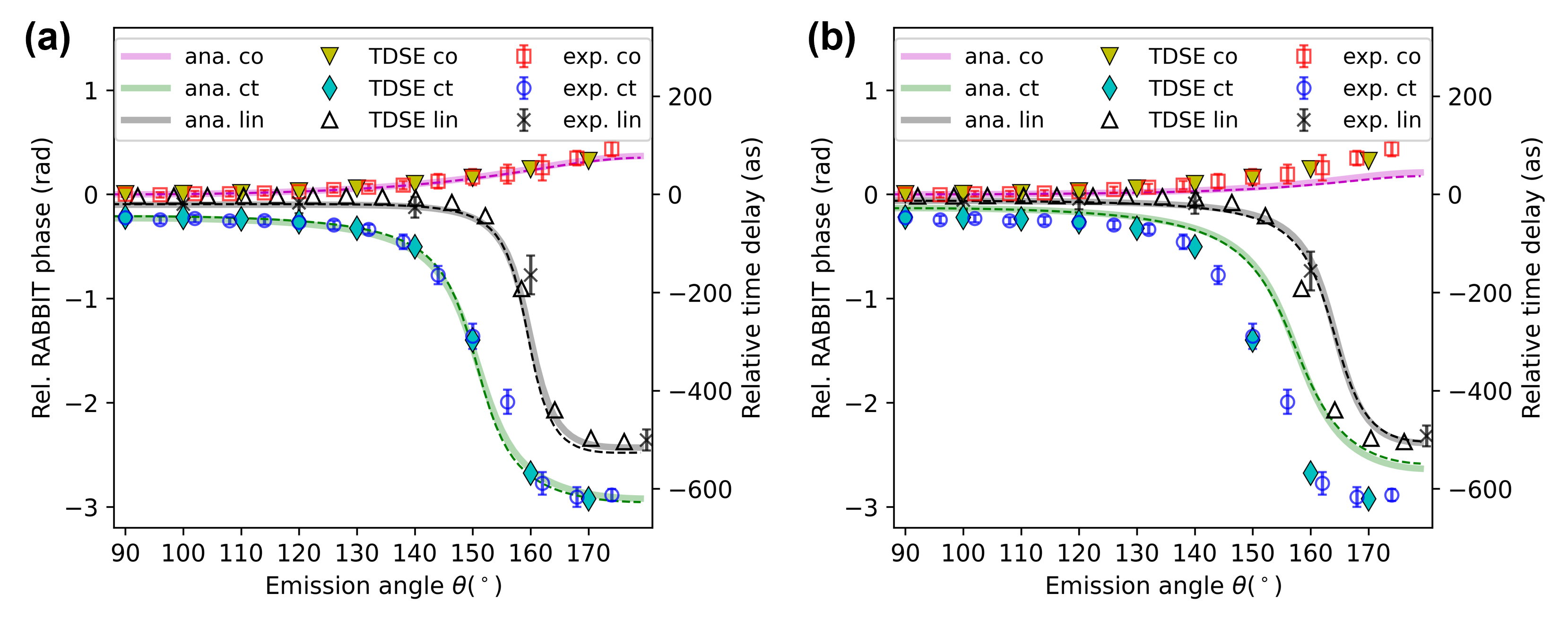}
\caption{Angle-dependent RABBIT phases and corresponding relative time delays of helium for co-rotating (co), counter-rotating (ct) and linear-linear (lin) XUV and dressing IR (800 nm) with final-state kinetic energy of 3.3 eV (SB18). The phases and relative time delays are referenced to the co-rotating case with electron emitted at $90^\circ$. For the co-rotating and counter-rotating cases, the experimental (exp.) and TDSE values are reported in \cite{han2024separation}. For the linear-linear case, the experimental values are reported in \cite{jiang2022atomic}, while the TDSE values are from \cite{kheifets2024characterization}. The one-photon ionisation amplitude is calculated using \texttt{ePolyScat} \cite{gianturco94a,natalense99a}. (a) The analytical (ana.) results are obtained from (\ref{eq:T_CC_in_F1}), including the calculation using unshifted and shifted center of expansion by 1 {\r A}, shown as the solid and dashed lines, respectively. (b) The results labelled by ``Boll'' are calculated using the formula reported in \cite{boll2022analytical}, including the calculation using unshifted and shifted center of expansion by 1 {\r A}, shown as the solid and dashed lines, respectively. \label{fig:He_circir_SB18_Appell}}
\end{center}
\end{figure}

An experimental test for the angular-momentum-dependent CC transition amplitude are angle-resolved time-delay measurements on atoms \cite{Bray2018,fuchs2020time,jiang2022atomic}. For example, using the (isotropic) asymptotic wavefunction for the intermediate state and the Kummer's function for the final state \cite{boll2022analytical}, Boll \textit{et al.} showed that this analytical formula with partial approximation matches well with the linear-XUV-linear-IR RABBIT experiments regarding the angle-dependent phase \cite{boll2023two,jiang2022atomic}. 
Here we additionally compare the theories with a circular-XUV-circular-IR RABBIT experiment reported recently by the present authors \cite{han2024separation}. The scheme for the generation and characterisation of the circular XUV attosecond pulse trains and the experimental setup for angle-resolved RABBIT have been described in our recent publications \cite{han2024separation,han2023attosecond,han2023optica}. The benefit of the circular-XUV-circular-IR RABBIT scheme for helium is that the difference in the co- and counter-rotating XUV and IR provides additional information of the partial waves. 
For example, if the circular XUV excites one of the helium electrons from $1s$ to the $E p_1$ intermediate state (the $z$-axis is defined as the light-propagation direction), in the CC transition step, if the dressing field is co-rotating with the XUV, then the absorption pathway leads only to the $E' d_2$ final state due to the restriction of the magnetic quantum number, while the emission pathway leads to the superposition of the $E' s$ and $E' d_0$ states; on the contrary, if the dressing field is counter-rotating, then the emission pathway yields the $E' d_2$ final state, and the absorption pathway gives the superposition of the $E' s$ and $E' d_0$ states. 
Since there are only three partial waves contributing to a given sideband, their individual moduli and phases can be extracted by the global fitting of the two-dimensional interference pattern \cite{villeneuve2017coherent} with only six parameters.
By comparing the co-rotating and counter-rotating cases, one can separate the Wigner part $1s \rightarrow E p$ and the continuum-continuum part $E p \rightarrow E' s/d$ of the photoionsation time delays for each final state \cite{han2024separation}. 
The angular dependence of the RABBIT phase is entirely absent from the isotropic asymptotic approximations, and the relative phase and modulus ratio between $E p \rightarrow E' s$ and $E p \rightarrow E' d$ play the key role, as explained in \cite{busto2019fano}. 
The theoretical and experimental results are compared in figure \ref{fig:He_circir_SB18_Appell} (a) and (b) using the analytical formulae from this work (\ref{eq:T_CC_in_F1}) and from Boll \textit{et al.} \cite{boll2022analytical}, respectively. We choose SB18 with relatively low electron kinetic energy (3.3 eV), where the asymptotic formulae have poor accuracy and are not shown in the figure. Our proposed analytical formula (\ref{eq:T_CC_in_F1}) shows remarkably good agreement with the TDSE and experimental values for both the co- and counter-rotating cases, as well as the linear-linear case, which is expected from the level of accuracy that it has shown in describing the moduli and phases of the transition amplitudes between different $l'$-states. The formula proposed by Boll \textit{et al.} is known to have less accuracy at lower electron kinetic energies and agrees with the TDSE semi-quantitatively, where the discrepancy in the experimental values is clearly observed in the counter-rotating case. Considering the precision of the experimental values, we conclude that our analytical formula quantitatively reproduces the experimental results and has comparable accuracy with the TDSE computation, which outperforms the formula proposed by Boll \textit{et al.} \cite{boll2022analytical} at low electron kinetic energies.

\begin{table}[h]
\caption{Partial-wave fractions of helium one-photon ionisation with expansion center shifted by 1 {\r A}, calculated using \texttt{ePolyScat} \cite{gianturco94a,natalense99a}.}
\label{tab:helium_l_components}
\begin{center}
\begin{tabular}{ccccccc}
\br
\textrm{$E_{\rm k}$ (eV)}&
\textrm{$l=0$}&
\textrm{$l=1$}&
\textrm{$l=2$}&
\textrm{$l=3$}&
\textrm{$l>3$}&\\
\mr
3.3  & 0.0808 & 0.7484 & 0.1617 & 0.0088 & 0.0002\\
18.8 & 0.1891 & 0.2266 & 0.3970 & 0.1599 & 0.0274\\
\br
\end{tabular}
\end{center}
\end{table}

\begin{figure}[bt]
\begin{center}
\includegraphics[width=1\textwidth]{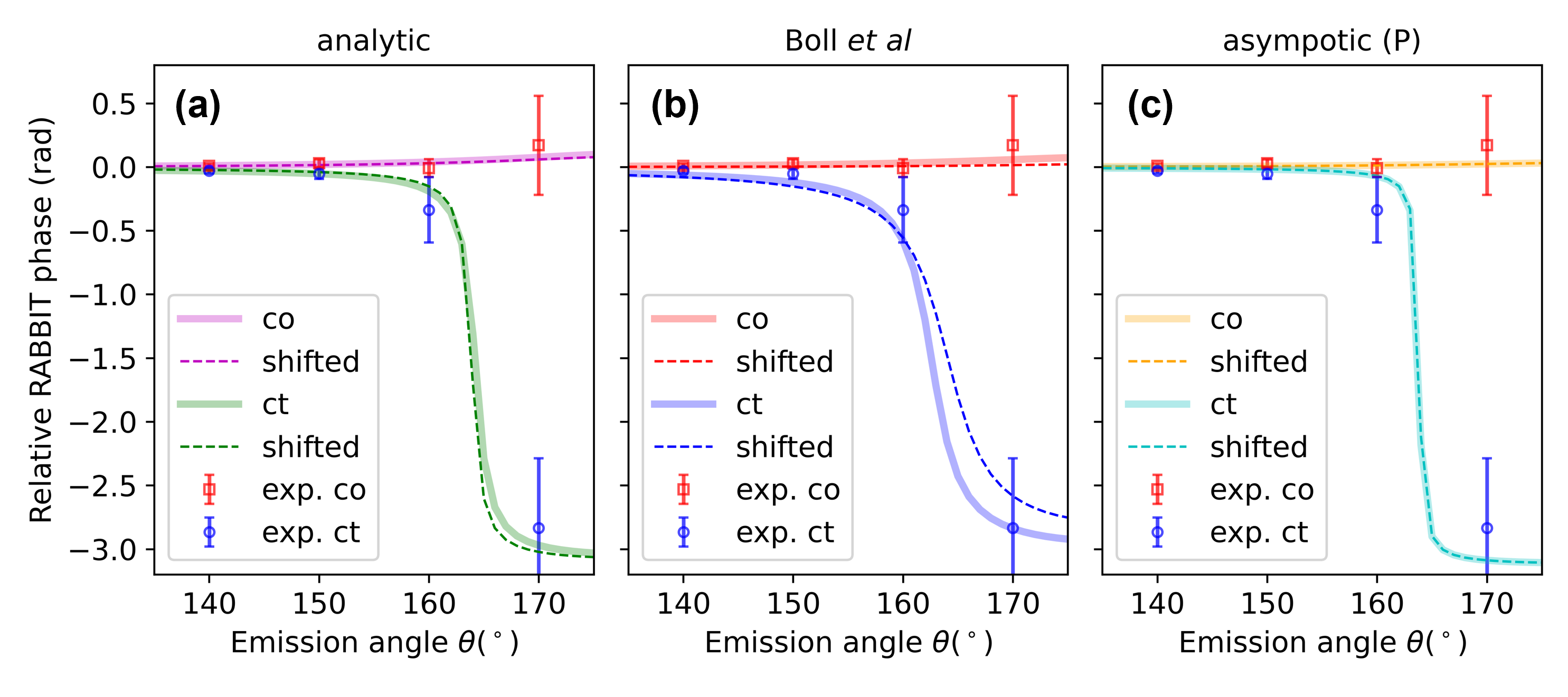}
\caption{Angle-dependent RABBIT phases of helium for co-rotating (co) and counter-rotating (ct) XUV and dressing IR (800 nm) with final-state energy of 18.8 eV (SB28). The phases and relative time delays are referenced to the co-rotating case with electron emitted at $90^\circ$. Panel (a) uses the analytic formula in this work (\ref{eq:T_CC_in_F1}); panel (b) uses the formula proposed by Boll \textit{et al.} \cite{boll2022analytical}; panel (c) uses the modified asymptotic formula ((P), equation (\ref{eq:T_CC_in_gamma}), $r_0=0$). In all panels, the experimental (exp.) values are taken from \cite{han2024separation}. The dashed lines with corresponding colours refer to the results of shifting the expansion center by 1 {\r A}. The one-photon ionisation amplitude is calculated using \texttt{ePolyScat} \cite{gianturco94a,natalense99a}. \label{fig:He_circir_SB28_compare}}
\end{center}
\end{figure}

In order to further check the sensitivity of our method regarding the shape of the potential, we performed calculations with helium one-photon photoionisation amplitudes with an expansion center displaced by 1 {\r A} using \texttt{ePolyScat} \cite{gianturco94a,natalense99a}. Since \texttt{ePolyScat} uses single-center expansions, when the atomic orbitals are displaced from the center of expansion, a series of (in principle infinite number of) partial waves are involved, in contrast to the centered helium atom, where the one-photon ionisation solely leads to the $p$-wave. The fractions (defined as the modulus square ratio) of partial waves of the shifted helium atom with electron kinetic energies of 3.3 eV and 18.8 eV are summarised in table \ref{tab:helium_l_components}. At 3.3 eV, $l=1$ ($p$-wave) is still dominating, with some contributions from $l=0$ and $l=2$ partial waves. As the electron kinetic energy increases to 18.8 eV, the distribution becomes more diffuse, with $l=2$ as the leading partial wave and comparable contributions from $l=0,1,3$. 
Although all the of observables at asymptotic distance from the origin should be insensitive to the shift of the atom, depending on the model, this may introduce artefacts. When the atom is displaced from the origin in the spherical coordinate, the potential is no longer central, and the unshifted Coulomb waves are no longer its eigenfunctions. On the other hand, one may argue that since the CC transition involves the outgoing electron wavepackets that escape from the atom, the transition amplitude should be less sensitive to the short-range potential, and the asymptotic behaviour is the same. The results using the analytical formula for the lower kinetic energy (3.3 eV) are shown as the dashed lines in figure \ref{fig:He_circir_SB18_Appell}, where the phase difference compared to the unshifted case is only marginal using both formulae from this work and from \cite{boll2022analytical}. For the higher kinetic energy (18.8 eV), we compare the results using the analytical formula (\ref{eq:T_CC_in_F1}), the formula proposed by Boll \textit{et al.} \cite{boll2022analytical}, and the modified asymptotic formula ((P), equation (\ref{eq:T_CC_in_gamma}), $r_0=0$) in figure \ref{fig:He_circir_SB28_compare}. Since the electron kinetic energy is relatively high, all three models show reasonable agreement with the experimental values. For the shifted helium, the time delay from the modified asymptotic formula is practically unchanged, since it is by its nature only sensitive to the asymptotic behaviour, whereas the formula proposed by Boll \textit{et al.} shows a non-negligible phase difference, indicating that the time delay may be less accurate when the potential deviates from the perfect central potential. The analytical formula shows much smaller difference, which implies that although taking the confluent hypergeometric function of the second kind may deviate from the reality in a non-central potential field, e.g. the potential of a molecular cation, the calculated time delay is largely converged. The application to molecular systems and comparison to the experimental results are in progress and will be published in a following work.

\subsection{Cooper-like minima for high-angular-momentum transitions}
\label{sec:high_L}

\begin{figure}[bt]
\begin{center}
\includegraphics[width=1\textwidth]{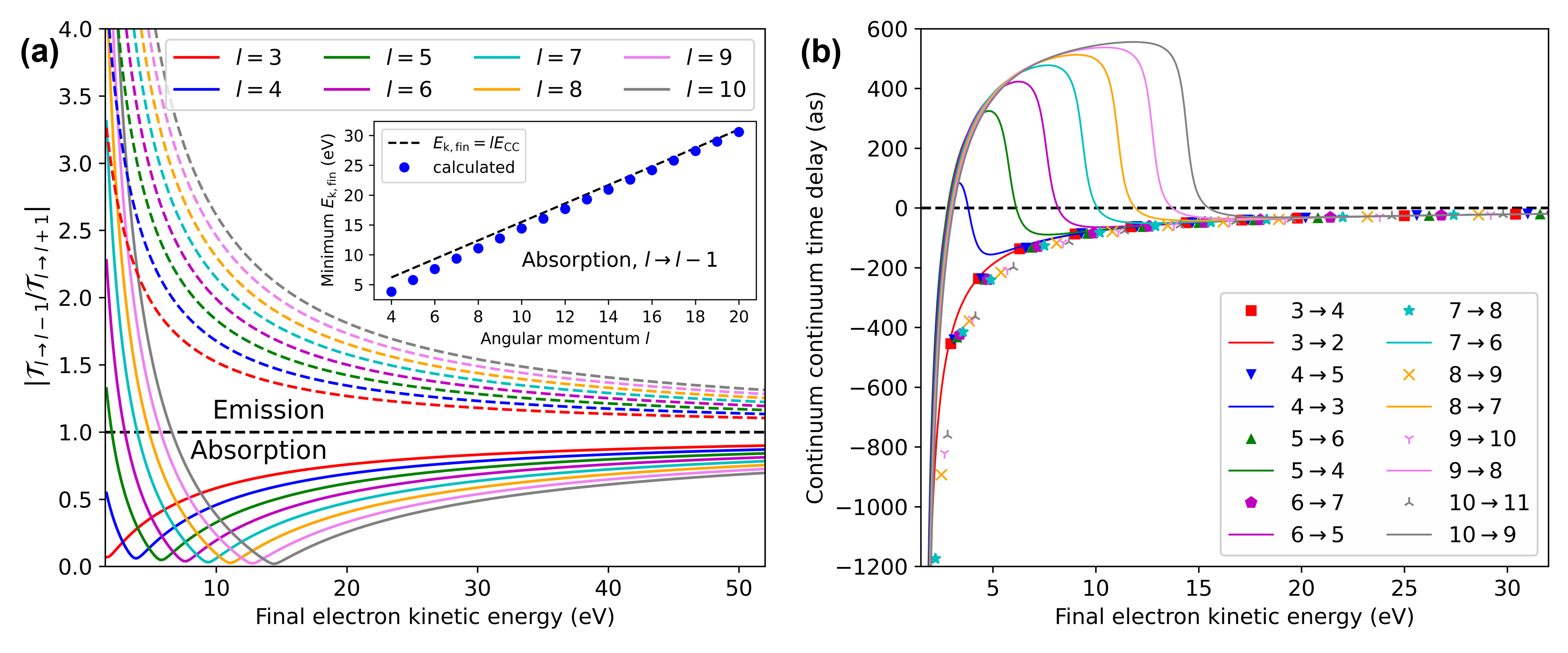}
\caption{(a) Moduli ratios for the absorption (solid) and emission (dashed) pathways (with dressing field wavelength of 800 nm) using the analytical formula (\ref{eq:T_CC_in_F1}) with higher angular momenta $l$s for the intermediate states. The inner panel shows the final-state electron kinetic energies at the minima of the absorption pathway of the $l \rightarrow l-1$ channels, in comparison of $l E_{\CC}$, where $E_{\CC}$ is the photon energy of the dressing field (1.55 eV). (b) The CC time delays defined by equation (\ref{eq:T_CC_finite_diff_def}) of various channels labelled as ``(intermediate angular momentum) $\rightarrow$ (final angular momentum)''. \label{fig:HighL_compare}}
\end{center}
\end{figure}

The analytical formula (\ref{eq:T_CC_in_F1}) can be extended into high $l$-states. Interestingly, we have observed Cooper-like minima \cite{cooper62a} for the absorption pathway of the $l \rightarrow l-1$ channels when $l$ is high enough, as illustrated in figure \ref{fig:HighL_compare} (a). The positions of the minima are approximately linear to the angular momentum $l$ of the intermediate state and approach $l E_{\CC}$ for large $l$s. For large angular momenta and low kinetic energies, the Fano’s propensity rule in CC transitions can be violated, namely $|\mathcal{T}^{\abs}_{l \rightarrow l-1} / \mathcal{T}^{\emi}_{l \rightarrow l-1}| > 1$. The detailed mathematical analysis, including the estimation of the positions of the minima using the asymptotic formula, can be found in \ref{sec:CM_explanation}. The corresponding time delays, as plotted in figure \ref{fig:HighL_compare} (b), are also structured near the minima and deviate from the general trend observed in figure \ref{fig:TransitonDelayCombined} and reported in \cite{Busto2024Atomic}, while the $l \rightarrow l+1$ channels do not have such feature and follow the same trend of $l \leq 3$.
Time delays near Cooper minima \cite{schoun14a,magrakvelidze2015attosecond,alexandridi2021attosecond} are known to have either local maximum or minimum, and in a work recently published by the present authors \cite{ji2024relation}, we demonstrate that the sign of the time delay relies on the topological property of the transition amplitude on the complex plane. Here, the Cooper-like minimum leads to a local maximum of the CC time delay, provided that the channel is pure. When the channel is coupled to other channels, as analysed in \cite{ji2024relation}, the time-delay structure may be affected.
Numerical verification and applications to Rydberg atoms and molecules are in progress and will be published as a following work.

\section{Conclusion}
\label{sec:Conclusion}

In conclusion, we have proposed an analytical formula for calculating the CC transition amplitudes for the hydrogen-like atoms which uses the confluent hypergeometric functions of the first and second kinds as the final and intermediate states, respectively. 
We first derived the integral between the confluent hypergeometric functions of the first and second kind, which can be expressed using the Appell's $F_1$ function.
Then we showed that this formula excellently reproduces the phase and the Fano's propensity rule in the CC transition, compared to the TDSE and other theoretical approaches.
In addition, we modified the WKB asymptotic approximation by expanding the terms to $r^{-1}$, to which the centrifugal potential contributes. The Fano's propensity rule and the main feature of the phases in CC transitions can be well captured with this relatively simple expression that involves only the gamma function.
The CC transition time delays derived from this analytical method coincide with the formula describing the Coulomb-laser coupling time delay, indicating that our approach may be extended for calculating the angle-dependent time delay in the streaking experiments. Detailed investigation in the streaking context is in progress and will be published in a separate work.
Finally, we used our method to calculate the relative phases of linear-linear and circular-circular RABBIT schemes at different electron emission angles. Even at kinetic energy as low as 3.3~eV, our proposed analytical formula quantitatively agrees with both the TDSE calculation and the experimental values.
We also checked the case with the shifted expansion center for calculating the spherical harmonics, and the results indicates that our method is not very sensitive to the shape of the short-range potential field, which opens the possibility of adapting our method to molecules.
For states with high angular momenta, our analytical formula predicts that the absorption pathway of $l \rightarrow l-1$ has a Cooper-like minimum, which affects its CC time delay. The detailed investigation will be presented in a following work.
We further note that the analytical form for the radial integral is not limited to the dipole transition, and the expression for multipole transitions is given in \ref{sec:multipole}. Liao \textit{et al.} recently investigated the linearly polarised RABBIT beyond the dipole approximation based on a modified version of the formula reported by Dahlstr{\"o}m \textit{et al.} \cite{liao2024linearly}. Our formula may also provide a more accurate evaluation of the non-dipole terms in the low-kinetic-energy region.


\section*{Acknowledgements}
We thank David Busto, Mathieu Gisselbrecht and Anatoli Kheifets for fruitful discussions. This work was funded by ETH Zürich. J.-B. Ji acknowledges the funding from the ETH grant 41-20-2. M. H.'s work was funded by the European Union’s Horizon 2020 research and innovation program under Marie Skłodowska-Curie agreement grant No. 801459, FP-RESOMUS.

\appendix

\section{Derivation for the radial integral}
\label{sec:Appendix_integral}

In order to obtain $\mathcal{J}^{0}_{\rho} (a,b,\lambda;n',\lambda';Q)$, we use the integration form of the confluent hypergeometric function of the second kind: 
\begin{equation}
    U(a,b;z) = \frac{1}{\Gamma(a)}
    \int_0^\infty \E^{-zt} t^{a-1} {(t+1)}^{b-a-1} \dd t
    \label{eq:U_int}
\end{equation}
and thus
\begin{eqnarray}
    \fl \mathcal{J}^{0}_{\rho} (a,b,\lambda;n',\lambda';Q)
    = \int_0^\infty \E^{-Q\xi} \xi^{\rho} 
    U(a,b;\lambda \xi) \Phi(-n',\rho+1;\lambda' \xi) \dd \xi \nonumber \\
    = \frac{1}{\Gamma(a)} \int_0^\infty \int_0^\infty
    \E^{-(Q + \lambda t) \xi} t^{a-1} {(t+1)}^{b-a-1}
    \xi^\rho \Phi(-n',\rho+1;\lambda' \xi) 
    \dd t \dd \xi .
\end{eqnarray}
By using the following formula for the ${_2F_1}$ function \cite{functionswolfram}, equation {(07.23.07.0003.01)}:
\begin{eqnarray}
     {_2F_1}(\alpha,\beta,\gamma;z)
    = \frac{1}{\Gamma(\beta)} 
    \int_0^\infty \E^{-\xi} \xi^{\beta-1}  \Phi(\alpha,\gamma;\xi z) \dd \xi
\end{eqnarray}
and swapping the order of integration, we have: 
\begin{eqnarray}
    \fl \mathcal{J}^{0}_{\rho} (a,b,\lambda;n',\lambda';Q) 
    = \frac{\rho!}{\Gamma(a)}
    \int_0^\infty 
    \frac{t^{a-1} {(t+1)}^{b-a-1}}{{(Q+\lambda t)}^{\rho+1}}
    {_2F_1} \left(-n',\rho+1,\rho+1,\frac{\lambda'}{Q+\lambda t} \right)
    \dd t .
\end{eqnarray}
Using the definition of the generalised hypergeometric functions \cite{functionswolfram}, equation {(07.19.02.0002.01)}:
\begin{equation}
    {_2F_1}(\alpha,\beta,\beta;z) = ~ _1F_0(\alpha;;z) = {(1-z)}^{-\alpha}
\end{equation}
we have: 
\begin{eqnarray}
    \fl \mathcal{J}^{0}_{\rho} (a,b,\lambda;n',\lambda';Q)
    = \frac{\rho!}{\Gamma(a)}
    \int_0^\infty 
    \frac{t^{a-1} {(t+1)}^{b-a-1}}{{(Q+\lambda t)}^{\rho+1}}  
    {\left( 1 - \frac{\lambda'}{Q+\lambda t} \right)}^{n'}
    \dd t \nonumber \\
    = \frac{\rho! ~ {(Q-\lambda')}^{n'}}{Q^{\rho+1+n'}}
    \int_0^\infty t^{a-1} {(t+1)}^{b-a-1}
    {\left( \frac{\lambda}{Q} t + 1 \right)}^{-(\rho+1+n')}
    {\left( \frac{\lambda}{Q-\lambda'} t + 1 \right)}^{n'} \dd t.
    \label{eq:J_0_int}
\end{eqnarray}
Let us use the integral representation of the Appell's $F_1$ function \cite{functionswolfram}, equation {(07.36.07.0001.01)}:
\begin{eqnarray}
    \fl F_1(\alpha;\beta_1,\beta_2;\gamma;z_1,z_2)
    = \frac{\Gamma(\gamma)}{\Gamma(\alpha) \Gamma(\gamma-\alpha)}
    \int_0^1 \frac{ h^{\alpha-1} (1-h)^{\gamma-\alpha-1} }{ (1 - h z_1)^{\beta_1} (1 - h z_2)^{\beta_2} } \dd h
    \nonumber \\
    = \frac{\Gamma(\gamma)}{\Gamma(\alpha) \Gamma(\gamma-\alpha)}
    \int_0^\infty \frac{ (t+1)^{\beta_1 + \beta_2 - \gamma}
    t^{\alpha-1} }{ {\left[ (1-z_1)t + 1 \right]}^{\beta_1}{\left[ (1-z_2)t + 1 \right]}^{\beta_2} } \dd t
\end{eqnarray}
where we used $h = \frac{t}{t+1}$. Compared to equation (\ref{eq:J_0_int}), we have $\alpha = a$, $\beta_1 = \rho+1+n'$, $\beta_2 = -n'$, $\gamma = \rho-b+a+2$, $z_1 = 1-\frac{\lambda}{Q}$, and $z_2 = 1 - \frac{\lambda}{Q-\lambda'}$. Therefore, we finally get equation (\ref{eq:J_0_rho_Q_in_F1}) in the main text.

\section{Formula for arbitrary $Z>0$}
\label{sec:multi_Z}

For the general case of $Z>0$ instead of $Z=1$, equation (\ref{eq:f_l}) becomes
\begin{eqnarray}
    \fl f_l(k,r;Z) = |C_l(k;Z)| ~ r^l ~ 
    \exp \left( {\I k r} \right) 
    \Phi \left( l+1-\I Z/k, 2l+2, -2\I k r \right)
    \label{eq:f_l_multi_Z}
\end{eqnarray}
where
\begin{equation}
    C_l(k;Z) = {k}^{l+1} ~ {2}^l ~ \E^{{\pi Z}/{(2 k)}} ~ 
    \frac{\Gamma \left( l+1+\I Z/k \right)}{(2l+1)!} ~ .
    \label{eq:C_k_l_multi_Z}
\end{equation}
Similarly, equation (\ref{eq:u_l}) becomes
\begin{eqnarray}
    \fl u_l(k,r;Z) = B_l(k;Z) ~ |C_l(k;Z)| ~ r^l ~ 
    \exp \left( {\I k r} \right)
    U \left( l+1-\I Z/k, 2l+2, -2\I k r \right)
    \label{eq:u_l_multi_Z}
\end{eqnarray}
with
\begin{equation}
    B_l(k;Z) = -2 \I \E^{-{\pi Z}/{k}} {(-1)}^{l} \frac{(2l+1)!}{\Gamma\left( l+1+\I Z/k \right)} ~ .
\end{equation}
Through the same derivation, the radial integral (\ref{eq:T_CC_in_F1}) can be generalised into: 
\begin{eqnarray}
\label{eq:T_CC_in_F1_multi_Z}
\fl \mathcal{T}_{l \rightarrow l'}(k,k';Z) = -\pi N_k N_{k'} \I^{l-l'-1} B_l(k;Z) C_l(k;Z) C_{l'}^*(k';Z) \\
\times \lim_{Q_0 \rightarrow 0^+} \mathcal{J}^{l-l'+2}_{2l'+1} \left( l+1-\frac{\I Z}{k},2l+2,-2\I k;-\left( l'+1-\frac{\I Z}{k'} \right),-2\I k'; Q_0 -\I(k+k') \right) \nonumber
\end{eqnarray} 
and its value can be obtained using equations (\ref{eq:J_sigma}) and (\ref{eq:J_0_rho_Q_in_F1}).

\section{Formula for multipole transitions}
\label{sec:multipole}

If the transition term is $r^{\mathcal{N}}$ instead of $r$ in equation (\ref{eq:CC_T}), where $\mathcal{N}$ is a positive integer, then equation (\ref{eq:T_CC_in_F1_multi_Z}) can be further generalised as:
\begin{eqnarray}
\label{eq:T_CC_in_F1_multi_Z_multipole}
\fl \mathcal{T}^{\mathcal{N}}_{l \rightarrow l'}(k,k';Z) = -\pi N_k N_{k'} \I^{l-l'-1} B_l(k;Z) C_l(k;Z) C_{l'}^*(k';Z) \\
\times \lim_{Q_0 \rightarrow 0^+} \mathcal{J}^{l-l'+1+\mathcal{N}}_{2l'+1} \left( l+1-\frac{\I Z}{k},2l+2,-2\I k;-\left( l'+1-\frac{\I Z}{k'} \right),-2\I k'; Q_0 -\I(k+k') \right) \nonumber
\end{eqnarray} 
and its value can also be obtained using equations (\ref{eq:J_sigma}) and (\ref{eq:J_0_rho_Q_in_F1}). The requirement of $l-l'+1+\mathcal{N} \geq 0$ is guaranteed by the selection rule that $|l-l'| \leq \mathcal{N}$.

\section{Explanation of Cooper-like minima}
\label{sec:CM_explanation}
The Cooper-like minimum in the absorption pathway of the $l \rightarrow l-1$ channel of CC transition originates from the transition amplitude between the irregular part of the intermediate-state wavefunction and the final-state wavefunction. Let us consider the asymptotic behaviour of the function $u_l(k,r)$ in expression (\ref{eq:u_l_asym}). The irregular ($g_l(k,r)$) and the regular ($f_l(k,r)$) parts behave like the cosine and sine functions in the far field, respectively. Due to the factor ${(-\I)}^{l}$, the wave of two neighboring $l$s will have phase shift of $\pi/2$; i.e. the sine and cosine are swapped. 
Following that the intermediate state has both the regular and irregular parts while the final state has only the regular part, the regular-regular transition can be roughly expressed as 
\begin{equation}
\fl \int_0^\infty \sin(kr) \cos(k'r) r \dd r = \frac{1}{2} \int_0^\infty \sin((k+k')r) + \sin((k-k')r) \dd r
\end{equation}
while the irregular-regular transition reads: 
\begin{equation}
\fl \int_0^\infty \cos(kr) \cos(k'r) r \dd r = \frac{1}{2} \int_0^\infty \cos((k+k')r) + \cos((k-k')r) \dd r ~ .
\end{equation}
These integrals can be evaluated using the following equations: 
\begin{equation}
    \int_0^\infty \sin(\mathfrak{K}r) \E^{-\lambda r} ~ r ~ \dd r = \frac{1}{\mathfrak{K}^2} \frac{2 \lambda / \mathfrak{K}}{{\left( {(\lambda/\mathfrak{K})}^2 + 1 \right)}^2}
    \label{eq:int_sin_exp_r}
\end{equation}
\begin{equation}
    \int_0^\infty \cos(\mathfrak{K}r) \E^{-\lambda r} ~ r ~ \dd r = \frac{1}{\mathfrak{K}^2} \frac{{(\lambda/\mathfrak{K})}^2 - 1}{{\left( {(\lambda/\mathfrak{K})}^2 + 1 \right)}^2}
    \label{eq:int_cos_exp_r}
\end{equation}
where $\mathfrak{K} = k \pm k'$. By taking the limit $\lambda \rightarrow 0^+$, (\ref{eq:int_sin_exp_r}) and (\ref{eq:int_cos_exp_r}) approach 0 and $-1/\mathfrak{K}^2$, respectively, and in the latter integral $\mathfrak{K} = k - k'$ contributes significantly more than $\mathfrak{K} = k + k'$, provided that the dressing-field photon energy is smaller than the final-state electron kinetic energy. Therefore, we have the integrals
\begin{equation}
    \mathcal{I}^{\mathrm{reg \rightarrow reg}}_{l \rightarrow l \pm 1} \sim 0 , 
\end{equation}
and
\begin{equation}
    \mathcal{I}^{\mathrm{reg \rightarrow irr}}_{l \rightarrow l \pm 1} \propto -\frac{1}{2 {\Delta k}^2} , 
    \label{eq:int_reg_irr_asym_rough}
\end{equation}
where $\Delta k \equiv k - k' = (k^2 - {k'}^2)/(k + k') = \Delta E/\bar{k} \approx \Delta E / {(2 \bar{E})}^{1/2} $, $\Delta E \equiv E - E' = \mp E_{\CC}$ for the absorption $(-)$ and emission $(+)$ pathways, $\bar{k} \equiv (k + k')/2$ stands for the averaged momentum of the intermediate and final states, and $\bar{E} \equiv (E + E')/2$ is the averaged electron kinetic energy. 
On the other hand, however, the assumption that the wavefunctions have the oscillating feature is only true, according to the WKB approximation, when the classical local kinetic energy is positive, which means 
\begin{equation}
    k^2 + \frac{2}{r} - \frac{l(l+1)}{r^2} = 0
\end{equation}
according to equation (\ref{eq:radial_equation}) under $Z=1$. When $l$ is large and $r$ is relatively small, the classical turning point is $\sim \sqrt{l(l+1)}/k \sim l/k$, which leads to the plateau for the $l \rightarrow l-1$ transition shown in figure \ref{fig:RadialWavefunction} (b) (red curve) between 0 and the classical turning point, i.e. inside the centrifugal barrier, where the integral in this particular part is
\begin{equation}
    \mathcal{I}^{\mathrm{barrier}}_{l \rightarrow l-1} \propto \int_0^{l/k} r \dd r = \frac{l^2}{2 k^2} .
\end{equation}
Note that the integral in (\ref{eq:int_cos_exp_r}) also has contribution inside the barrier, but there is a factor of $1/2$ so it contributes less than the exact formula, and overall the exact value is added by a positive component $\zeta^2 l^2/(2 k^2)$, where $\zeta>0$. Using (\ref{eq:int_reg_irr_asym_rough}), the total integral crosses zero when $\zeta^2 l^2/(2 k^2) = 1/({2 \Delta k}^2) = {\bar{k}}^2/(2 {\Delta E}^2)$, namely $k^2 \approx \zeta l E_{\CC}$, which gives the observed linear scaling of the positions of the minima. 
On the contrary, as shown in figure \ref{fig:RadialWavefunction} (c), for the $l \rightarrow l+1$ transition, the product of the irregular part of the intermediate state and the final state near the origin approaches zero, which leads to a negative term added to the overall integral and avoids the zero point.

This can also be semi-quantitatively derived from the modified asymptotic formula (\ref{eq:T_CC_in_gamma}). Letting $Z=1$ and assuming $k$ is relatively large, we use the rough approximation of $\Gamma(2+\I/k-\I/k') \approx \Gamma(2) = 1$ and $\Gamma(1+\I/k-\I/k') \approx \Gamma(1) = 1$, and $|{[\I(k'-k)]}^{2+\I/k-\I/k'}| \approx {\Delta k}^2$, so the zero of $|\mathcal{T}_{l \rightarrow l'}(k,k')|$ is reached when $1 + (k-k')(q-q') = 0$ in formula (\ref{eq:T_CC_in_gamma}). For large $l$, $q \approx b / k = l(l+1)/(2k)$ and $q' \approx b' / k' = l(l-1)/(2k')$, where we have used $l'= l-1$. Therefore, $ 2 k k' (q - q') = l^2 (k' - k) + l (k' + k) $, and the condition for the zero becomes: 
\begin{eqnarray}
    \fl 0  = 1 - \frac{E_{\CC}}{\bar{k}}
    \left( \frac{E_{\CC}}{\bar{k}} \frac{l^2}{2 k k'} + \frac{l \bar{k}}{k k'} \right)
     = 1 - \frac{l E_{\CC}}{k k'} - \frac{l^2 {E_{\CC}}^2}{2 {\bar{k}}^2 k k'} 
     \approx 1 - \frac{l E_{\CC}}{{\bar{k}}^2} - \frac{1}{2} \frac{l^2 {E_{\CC}}^2}{{\bar{k}}^4} .
    \label{eq:equation_gamma_zero_approx}
\end{eqnarray}
Solving this equation yields
\begin{equation}
    \bar{E} = {\bar{k}}^2 / 2 = \left( \frac{1+\sqrt{3}}{4} \right) l E_{\CC} .
\end{equation}
Although this coefficient ($\sim 0.7$) is smaller than the observation ($\sim 1.0$), considering the level of approximation, this has reflected the main feature of the behaviour of the transition amplitude between the irregular part of the intermediate state with $l$ and the final state with $l-1$ by absorbing one photon from the dressing field. 




\section*{References}
\bibliographystyle{iopart-num}
\bibliography{attobib,CCAppell}

\providecommand{\newblock}{}
\begin{thebibliography}{10}
\expandafter\ifx\csname url\endcsname\relax
  \def\url#1{{\tt #1}}\fi
\expandafter\ifx\csname urlprefix\endcsname\relax\def\urlprefix{URL }\fi
\providecommand{\eprint}[2][]{\url{#2}}

\bibitem{nobelprizeNobelPrize}
{Nobel Prize Outreach AB 2024} 2024 {T}he {N}obel {P}rize in {P}hysics 2023. {N}obelprize.org \url{https://www.nobelprize.org/prizes/physics/2023/summary/} [Accessed 05-07-2024]

\bibitem{mcpherson87a}
McPherson A, Gibson G, Jara H, Johann U, Luk T~S, McIntyre I~A, Boyer K and Rhodes C~K 1987 {\em JOSA B\/} {\bf 4} 595

\bibitem{ferray88a}
Ferray M, L'Huillier A, Li X~F, Lompre L~A, Mainfray G and Manus C 1988 {\em J. Phys. B\/} {\bf 21} L31

\bibitem{itatani02a}
Itatani J, Qu{\'e}r{\'e} F, Yudin G~L, Ivanov M~Y, Krausz F and Corkum P~B 2002 {\em Phys. Rev. Lett.\/} {\bf 88} 173903

\bibitem{kienberger03a}
Kienberger R, Goulielmakis E, Uiberacker M, Baltuska A, Yakovlev V, Bammer F, Scrinzi A, Westerwalbesloh T, Kleineberg U, Heinzmann U, Drescher M and Krausz F 2003 {\em Nature\/} {\bf 427} 817--821

\bibitem{gaumnitz17a}
Gaumnitz T, Jain A, Pertot Y, Huppert M, Jordan I, Ardana-Lamas F and W\"{o}rner H~J 2017 {\em Optics Express\/} {\bf 25} 27506--27518

\bibitem{paul01a}
Paul P~M, Toma E~S, Breger P, Mullot G, Aug\'e F, Balcou P, Muller H~G and Agostini P 2001 {\em Science\/} {\bf 292} 1689

\bibitem{muller02a}
Muller H 2002 {\em Applied Physics B\/} {\bf 74} s17--s21 ISSN 1432-0649 \urlprefix\url{https://doi.org/10.1007/s00340-002-0894-8}

\bibitem{dorner00a}
D\"{o}rner R, Mergel V, Jagutzki O, Spielberger L, Ullrich J, Moshammer R and Schmidt-B\"{o}cking H 2000 {\em Physics Reports\/} {\bf 330} 95--192

\bibitem{ullrich03a}
Ullrich J, Moshammer R, Dorn A, D\"{o}rner R, Schmidt L and Schmidt-B\"{o}cking H 2003 {\em Rep. Prog. Phys.\/} {\bf 66} 1463

\bibitem{gong2022attosecond}
Gong X, Heck S, Jelovina D, Perry C, Zinchenko K, Lucchese R and W{\"o}rner H~J 2022 {\em Nature\/} {\bf 609} 507--511

\bibitem{eisenbud1948formal}
Eisenbud L 1948 {\em The formal properties of nuclear collisions\/} (Princeton University)

\bibitem{wigner55a}
Wigner E~P 1955 {\em Phys Rev A\/} {\bf 98} 145

\bibitem{smith60a}
Smith F~T 1960 {\em Phys. Rev.\/} {\bf 118} 349--356 \urlprefix\url{http://link.aps.org/doi/10.1103/PhysRev.118.349}

\bibitem{amusia1972interference}
Amusia M~Y, Ivanov V, Cherepkov N and Chernysheva L 1972 {\em Physics Letters A\/} {\bf 40} 361--362

\bibitem{kheifets13a}
Kheifets A~S 2013 {\em Phys. Rev. A\/} {\bf 87} 063404-- \urlprefix\url{http://link.aps.org/doi/10.1103/PhysRevA.87.063404}

\bibitem{zangwill1980density}
Zangwill A and Soven P 1980 {\em Physical Review A\/} {\bf 21} 1561

\bibitem{dixit2013time}
Dixit G, Chakraborty H~S and Madjet M~E~A 2013 {\em Physical review letters\/} {\bf 111} 203003

\bibitem{gianturco94a}
Gianturco F~A, Lucchese R~R and Sanna N 1994 {\em J. Chem. Phys.\/} {\bf 100} 6464--6471 \urlprefix\url{http://link.aip.org/link/?JCP/100/6464/1}

\bibitem{natalense99a}
Natalense A~P~P and Lucchese R~R 1999 {\em J. Chem. Phys.\/} {\bf 111} 5344--5348 \urlprefix\url{http://link.aip.org/link/?JCP/111/5344/1}

\bibitem{Busto2024Atomic}
Busto D, Zhong S, Dahlstr{\"o}m J~M, L'Huillier A and Gisselbrecht M 2024 {\em Ultrafast Electronic and Structural Dynamics\/} (Singapore: Springer Nature Singapore) chap~1, pp 1--43

\bibitem{dahlstrom13a}
Dahlstr\"om J, Gu\'enot D, Kl\"under K, Gisselbrecht M, Mauritsson J, L'Huillier A, Maquet A and Ta\"{i}eb R 2013 {\em Chemical Physics\/} {\bf 414} 53--64 ISSN 0301-0104 \urlprefix\url{http://www.sciencedirect.com/science/article/pii/S0301010412000298}

\bibitem{dahlstrom12a}
Dahlstr\"om J~M, L'Huillier A and Maquet A 2012 {\em Journal of Physics B: Atomic, Molecular and Optical Physics\/} {\bf 45} 183001 ISSN 0953-4075 \urlprefix\url{http://stacks.iop.org/0953-4075/45/i=18/a=183001}

\bibitem{marante2014hybrid}
Marante C, Argenti L and Mart{\'\i}n F 2014 {\em Physical Review A\/} {\bf 90} 012506

\bibitem{kluender11a}
Kl\"under K, Dahlstr\"om J~M, Gisselbrecht M, Fordell T, Swoboda M, Gu\'enot D, Johnsson P, Caillat J, Mauritsson J, Maquet A, Ta\"{i}eb and L'Huillier A 2011 {\em Physical Review Letters\/} {\bf 106}(14) 143002 \urlprefix\url{http://link.aps.org/doi/10.1103/PhysRevLett.106.143002}

\bibitem{guenot12a}
Gu\'enot D, Kl\"under K, Arnold C~L, Kroon D, Dahlstr\"om J~M, Miranda M, Fordell T, Gisselbrecht M, Johnsson P, Mauritsson J, Lindroth E, Maquet A, Ta\"{\i}eb R, L'Huillier A and Kheifets A~S 2012 {\em Phys. Rev. A\/} {\bf 85}(5) 053424 \urlprefix\url{http://link.aps.org/doi/10.1103/PhysRevA.85.053424}

\bibitem{huppert16a}
Huppert M, Jordan I, Baykusheva D, Von~Conta A and W{\"o}rner H~J 2016 {\em Physical Review Letters\/} {\bf 117} 093001

\bibitem{baykusheva17a}
Baykusheva D and W\"{o}rner H~J 2017 {\em The Journal of Chemical Physics\/} {\bf 146} 124306 ISSN 0021-9606 \urlprefix\url{http://dx.doi.org/10.1063/1.4977933}

\bibitem{isinger17a}
Isinger M, Squibb R, Busto D, Zhong S, Harth A, Kroon D, Nandi S, Arnold C, Miranda M, Dahlstr{\"o}m J~M {\em et~al.\/} 2017 {\em Science\/} {\bf 358} 893--896

\bibitem{heck2021attosecond}
Heck S, Baykusheva D, Han M, Ji J~B, Perry C, Gong X and W{\"o}rner H~J 2021 {\em Science Advances\/} {\bf 7} eabj8121

\bibitem{jordan20a}
Jordan I, Huppert M, Rattenbacher D, Peper M, Jelovina D, Perry C, von Conta A, Schild A and W{\"o}rner H~J 2020 {\em Science\/} {\bf 369} 974--979 ISSN 0036-8075 (\textit{Preprint} \eprint{https://science.sciencemag.org/content/369/6506/974.full.pdf}) \urlprefix\url{https://science.sciencemag.org/content/369/6506/974}

\bibitem{heck2022two}
Heck S, Han M, Jelovina D, Ji J~B, Perry C, Gong X, Lucchese R, Ueda K and W{\"o}rner H~J 2022 {\em Physical Review Letters\/} {\bf 129} 133002

\bibitem{heuser16a}
Heuser S, Jim\'enez~Gal\'an A, Cirelli C, Marante C, Sabbar M, Boge R, Lucchini M, Gallmann L, Ivanov I, Kheifets A~S, Dahlstr\"om J~M, Lindroth E, Argenti L, Mart\'{\i}n F and Keller U 2016 {\em Phys. Rev. A\/} {\bf 94}(6) 063409 \urlprefix\url{https://link.aps.org/doi/10.1103/PhysRevA.94.063409}

\bibitem{fuchs2020time}
Fuchs J, Douguet N, Donsa S, Martin F, Burgd{\"o}rfer J, Argenti L, Cattaneo L and Keller U 2020 {\em Optica\/} {\bf 7} 154--161

\bibitem{jiang2022atomic}
Jiang W, Armstrong G~S, Tong J, Xu Y, Zuo Z, Qiang J, Lu P, Clarke D~D, Benda J, Fleischer A {\em et~al.\/} 2022 {\em Nature Communications\/} {\bf 13} 5072

\bibitem{busto2019fano}
Busto D, Vinbladh J, Zhong S, Isinger M, Nandi S, Maclot S, Johnsson P, Gisselbrecht M, L’Huillier A, Lindroth E {\em et~al.\/} 2019 {\em Physical review letters\/} {\bf 123} 133201

\bibitem{han2024separation}
Han M, Ji J~B, Leung C~S, Ueda K and W{\"o}rner H~J 2024 {\em Science Advances\/} {\bf 10} eadj2629

\bibitem{boll2022analytical}
Boll D~I~R, Martini L and Fojon O~A 2022 {\em Physical Review A\/} {\bf 106} 023116

\bibitem{boll2023two}
Boll D~I~R, Martini L, Palacios A and Fojon O~A 2023 {\em Physical Review A\/} {\bf 107} 043113

\bibitem{aymar1980two}
Aymar M and Crance M 1980 {\em Journal of Physics B: Atomic and Molecular Physics\/} {\bf 13} L287

\bibitem{shakeshaft1986sturmian}
Shakeshaft R 1986 {\em Physical Review A\/} {\bf 34} 244

\bibitem{DLMF}
{\it NIST Digital Library of Mathematical Functions} \url{https://dlmf.nist.gov/}, Release 1.2.1 of 2024-06-15 f.~W.~J. Olver, A.~B. {Olde Daalhuis}, D.~W. Lozier, B.~I. Schneider, R.~F. Boisvert, C.~W. Clark, B.~R. Miller, B.~V. Saunders, H.~S. Cohl, and M.~A. McClain, eds. \urlprefix\url{https://dlmf.nist.gov/}

\bibitem{gaspard2018connection}
Gaspard D 2018 {\em Journal of Mathematical Physics\/} {\bf 59}

\bibitem{shakeshaft1985note}
Shakeshaft R 1985 {\em Journal of Physics B: Atomic and Molecular Physics\/} {\bf 18} L611

\bibitem{meixner1933greensche}
Meixner J 1933 {\em Mathematische Zeitschrift\/} {\bf 36} 677--707

\bibitem{hostler1964coulomb}
Hostler L 1964 {\em Journal of Mathematical Physics\/} {\bf 5} 591--611

\bibitem{hostler1970coulomb}
Hostler L~C 1970 {\em Journal of Mathematical Physics\/} {\bf 11} 2966--2970

\bibitem{benda2024angular}
Benda J, Ma{\v{s}}{\'\i}n Z, Palakkal S, L{\'e}pine F, Nandi S and Loriot V 2024 {\em arXiv preprint arXiv:2407.14160\/}

\bibitem{Zorich2016}
Zorich V~A 2016 {\em Integrals Depending on a Parameter\/} (Berlin, Heidelberg: Springer Berlin Heidelberg) pp 405--492 ISBN 978-3-662-48993-2 \urlprefix\url{https://doi.org/10.1007/978-3-662-48993-2_9}

\bibitem{gordon1929berechnung}
Gordon W 1929 {\em Annalen der Physik\/} {\bf 394} 1031--1056

\bibitem{tarasov1995multipole}
Tarasov V 1995 {\em International Journal of Modern Physics B\/} {\bf 9} 2699--2718

\bibitem{tarasov2003w}
Tarasov V 2003 {\em Journal of Mathematical Physics\/} {\bf 44} 1449--1452

\bibitem{saad2003integrals}
Saad N and Hall R~L 2003 {\em Journal of Physics A: Mathematical and General\/} {\bf 36} 7771

\bibitem{functionswolfram}
Wolfram~Research I 2024 The mathematical functions site \url{https://functions.wolfram.com/} (Accessed on 07/01/2024)

\bibitem{Bateman:100233}
Bateman H and Erdélyi A 1955 {\em {Higher transcendental functions}\/} ({\em California Institute of technology. Bateman Manuscript project\/} no~1) (New York, NY: McGraw-Hill) \urlprefix\url{https://cds.cern.ch/record/100233}

\bibitem{karule1978two}
Karule E 1978 {\em Journal of Physics B: Atomic and Molecular Physics\/} {\bf 11} 441

\bibitem{karule2003general}
Karule E and Moine B 2003 {\em Journal of Physics B: Atomic, Molecular and Optical Physics\/} {\bf 36} 1963

\bibitem{dubuc1990approximation}
Dubuc S 1990 {\em Journal of mathematical analysis and applications\/} {\bf 146} 461--468

\bibitem{dahlstrom12b}
Dahlstr{\"{o}}m J~M, Carette T and Lindroth E 2012 {\em Physical Review A - Atomic, Molecular, and Optical Physics\/} {\bf 86} 1--4 ISSN 10502947

\bibitem{berkane2024probing}
Berkane M, L{\'e}v{\^e}que C, Ta{\"\i}eb R, Caillat J and Dubois J 2024 {\em Physical Review A\/} {\bf 110} 013120

\bibitem{oliphant2007python}
Oliphant T~E 2007 {\em Computing in science \& engineering\/} {\bf 9} 10--20

\bibitem{harris2020array}
Harris C~R, Millman K~J, Van Der~Walt S~J, Gommers R, Virtanen P, Cournapeau D, Wieser E, Taylor J, Berg S, Smith N~J {\em et~al.\/} 2020 {\em Nature\/} {\bf 585} 357--362

\bibitem{mpmath}
{The mpmath development team} 2023 {\em mpmath: a {P}ython library for arbitrary-precision floating-point arithmetic (version 1.3.0)\/} {\tt http://mpmath.org/}

\bibitem{jayadevan2001two}
Jayadevan A and Thayyullathil R~B 2001 {\em Journal of Physics B: Atomic, Molecular and Optical Physics\/} {\bf 34} 699

\bibitem{serov2015interpretation}
Serov V~V, Derbov V~L and Sergeeva T~A 2015 {\em Interpretation of the time delay in the ionization of Coulomb systems by attosecond laser pulses\/} (Springer) pp 213--230

\bibitem{ivanov2011accurate}
Ivanov M and Smirnova O 2011 {\em Physical Review Letters\/} {\bf 107} 213605

\bibitem{kheifets2022ionization}
Kheifets A, Wielian R, Serov V, Ivanov I~A, Wang A~L, Marinelli A and Cryan J~P 2022 {\em Physical Review A\/} {\bf 106} 033106

\bibitem{kheifets2024characterization}
Kheifets A~S 2024 {\em Physical Review Research\/} {\bf 6} L012002

\bibitem{Bray2018}
Bray A~W, Naseem F and Kheifets A~S 2018 {\em Physical Review A\/} {\bf 97} 1--12 ISSN 24699934

\bibitem{han2023attosecond}
Han M, Ji J~B, Bal{\v{c}}i{\=u}nas T, Ueda K and W{\"o}rner H~J 2023 {\em Nature Physics\/} {\bf 19} 230--236

\bibitem{han2023optica}
Han M, Ji J~B, Ueda K and W{\"o}rner H~J 2023 {\em Optica\/} {\bf 10} 1044--1052

\bibitem{villeneuve2017coherent}
Villeneuve D, Hockett P, Vrakking M and Niikura H 2017 {\em Science\/} {\bf 356} 1150--1153

\bibitem{cooper62a}
Cooper J~W 1962 {\em Phys. Rev.\/} {\bf 128} 681

\bibitem{schoun14a}
Schoun S, Chirla R, Wheeler J, Roedig C, Agostini P, DiMauro L, Schafer K and Gaarde M 2014 {\em Physical Review Letters\/} {\bf 112} 153001

\bibitem{magrakvelidze2015attosecond}
Magrakvelidze M, Madjet M~E~A, Dixit G, Ivanov M and Chakraborty H~S 2015 {\em Physical Review A\/} {\bf 91} 063415

\bibitem{alexandridi2021attosecond}
Alexandridi C, Platzer D, Barreau L, Busto D, Zhong S, Turconi M, Neori{\v{c}}i{\'c} L, Laurell H, Arnold C, Borot A {\em et~al.\/} 2021 {\em Physical Review Research\/} {\bf 3} L012012

\bibitem{ji2024relation}
Ji J~B, Kheifets A~S, Han M, Ueda K and W{\"o}rner H~J 2024 {\em New Journal of Physics\/} {\bf 26} 093014

\bibitem{liao2024linearly}
Liao Y, Chen Y, Dahlstr{\"o}m J~M, Pi L~W, Lu P and Zhou Y 2024 {\em Physical Review A\/} {\bf 110} 023109

\end{thebibliography}

\end{document}